\documentclass[conference]{IEEEtran}
\IEEEoverridecommandlockouts
\usepackage{amssymb}
\usepackage{hyperref}
\usepackage{cite}
\usepackage{graphicx}
\usepackage{array}
\usepackage{multirow}
\usepackage{amsmath}
\usepackage{stfloats}
\usepackage{graphicx}
\usepackage{subfigure}
\usepackage{tabularx}
\usepackage{booktabs}
\usepackage{epsfig,epsf}
\usepackage{setspace}
\usepackage{bm}
\usepackage{textcomp}
\usepackage{algorithmic}
\usepackage{algorithm}
\usepackage{subfigure}
\usepackage{caption}
\usepackage{graphicx}
\usepackage{setspace}
\usepackage{url}
\usepackage{verbatim}
\usepackage{graphicx}
\usepackage{cite}
\usepackage{amssymb}
\usepackage{color}
\usepackage{xpatch}
\definecolor{shadecolor}{RGB}{0,0,255}
\definecolor{blue}{RGB}{0,0,255}

\newtheorem{theorem}{Theorem}
\newtheorem{lemma}{Lemma}
\newtheorem{proposition}{Proposition}

\newtheorem{remark}{Remark}
\usepackage{soul}
\usepackage{booktabs} 
\usepackage{tabularx}
\usepackage{array} 
\usepackage[table]{xcolor} 
\definecolor{tableblue}{RGB}{221,238,248} \newcolumntype{Y}{>{\raggedright\arraybackslash}X}

\makeatletter

\ExplSyntaxOn
\cs_new:Npn \bibColoredItems #1#2
{
	\clist_map_inline:nn {#2} { \cs_new:cpn {bib@colored@##1} {#1} } 
}
\ExplSyntaxOff

\newcommand\bib@setcolor[1]{%
	\ifcsname bib@colored@#1\endcsname
	\expanded{\noexpand\color{\csname bib@colored@#1\endcsname}}%
	\else
	\normalcolor
	\fi
}

\makeatother

\begin{document}
	
\title{Toward Time-Invariant Rydberg Atomic Quantum Receivers for AFDM Sensing}

\author{Qihao Peng,~\IEEEmembership{Member,~IEEE},
         Qu Luo,~\IEEEmembership{Member,~IEEE},
          Jizhou Wu, Tianqi Mao,~\IEEEmembership{Member,~IEEE}, \\
         Neng Ye,~\IEEEmembership{Senior Member,~IEEE}, Cunhua Pan,~\IEEEmembership{Senior Member,~IEEE},
       Pei Xiao,~\IEEEmembership{Senior Member,~IEEE},\\ Jiangzhou Wang,~\IEEEmembership{Fellow,~IEEE},
       
       \vspace{-2cm}
   
		\thanks{Q. Peng, Q. Luo, and P. Xiao are affiliated with 5G and 6G Innovation Centre, Institute for Communication Systems (ICS) of the University of Surrey, Guildford, GU2 7XH, UK. (e-mail: \{q.peng, q.u.luo,p.xiao\}@surrey.ac.uk). } \\
              \thanks{J. Wu is with State Key Laboratory of Quantum Optics Technologies and Devices, Institute of Laser Spectroscopy, Shanxi University, Taiyuan 030006, China (e-mail: wujz@sxu.edu.cn).} \\
          \thanks{T. Mao is with the State Key Laboratory of Environment Characteristics and Effects for Near-Space, Beijing Institute of Technology, Beijing 100081, China (e-mail: maotq@bit.edu.cn).}
        \thanks{N. Ye is the School of Cyberspace Science and Technology, Beijing Institute of Technology, Beijing 100081, China (email: ianye@bit.edu.cn).}\\
           \thanks{C. Pan and J. Wang are with the National Mobile Communications Research Laboratory, Southeast University, Nanjing, China (e-mail: cpan@seu.edu.cn;j.z.wang@seu.edu.cn).}  \\

        }

	
	\maketitle

\begin{abstract}
In this paper, we propose a five-level Rydberg atomic quantum receiver (RAQR) framework for affine frequency division multiplexing (AFDM) sensing, where wideband echoes are dechirped by a chirp-based local oscillator (LO). To mitigate the impacts of the LO-induced time-varying bias and conversion gain, an auxiliary (AUX) microwave field is introduced to compensate for the LO-induced light shift through a joint Stark-and-gain lock.  Under far detuning, we establish the equivalence to a dressed three-level electromagnetically induced transparency, enabling time-invariant atomic conversion, and then derive the closed-form expressions for the optimal operating point and LO dressing. Furthermore, the effects of LO and AUX driving on the
signal-to-noise ratio and Cram\'er--Rao bounds (CRBs) are analyzed, revealing a range-accuracy floor and a delay--Doppler tradeoff. Simulation results validate the analysis and confirm that the proposed framework maintains a time-invariant atomic response, facilitating efficient fast Fourier transform (FFT)-based sensing and providing a promising approach to AFDM-based RAQRs.
\end{abstract}


\section{Introduction}
The sixth-generation (6G) networks envision integrated sensing and communication (ISAC) as a native service, which demands receivers with wide bandwidth, high sensitivity, and multiband operation \cite{wang2023road}. However, conventional antenna-based receivers face fundamental limits in meeting these demands. Particularly, improving the sensitivity against front-end noise favors larger antenna apertures, whereas the Chu limit constrains the efficient bandwidth of electrically small passive antennas~\cite{chu1948physical}. Meanwhile, multi-band operation may require additional antennas and radio-frequency (RF) chains, increasing the hardware complexity. Therefore, it is desirable to explore the novel receiver based on fundamentally different sensing mechanisms for 6G ISAC \cite{11363373,11668904}.

Rydberg atomic quantum receivers (RAQRs) offer an alternative sensing mechanism by optically detecting RF fields through their interactions with Rydberg atoms \cite{Chen2025,fancher2021rydberg}. Probe and coupling lasers can establish electromagnetically induced transparency (EIT), and this window can be split by Autler–Townes splitting (ATS) when a resonant microwave is applied \cite{sedlacek2012microwave}. Then, with a local oscillator (LO) on the same transition, the amplitude and phase of a weak field are further mapped onto the transmitted probe power, which enables heterodyne quantum sensing \cite{jing2020atomic}. By converting the electromagnetic signal into an optical readout, the intrinsic noise of an RAQR is set by quantum projection and photon shot noise rather than electronic thermal noise, which enables the potential for high-sensitivity reception. Meanwhile, its electrically small
sensing volume is not constrained by the Chu limit \cite{cox2018quantum}, and the abundant Rydberg transitions allow SI-traceable detection over a broad frequency range \cite{simons2016using}.

Owing to the appealing properties, RAQRs have been extensively studied. For bridging the electromagnetic and optical domains, the magnitude-based models of RAQRs were established by \cite{CuiJSACatomc}, while the phase needs to be detected in an iterative manner. By leveraging the superheterodyne architecture, the equivalent complex baseband models were proposed by \cite{Gongtcom,11568941}. Considering different atomic excitation configurations, a unified analytical model was derived in \cite{xiang2026unifiedanalyticalmodelatomic}. 
To overcome the bandwidth limitations imposed by atomic response, wideband RAQRs have been designed based on a six-wave mixing approach \cite{chen2026widebandquantumtransductionrydberg}, a frequency-comb-based architecture \cite{11655192}, multi-band reception \cite{Cuimultiband}, and a Floquet-based 5-level configuration \cite{peng2026continuouswidebandmulticarrierrydberg}.
Beyond the equivalent model, the relationship between the input signal and noise was revealed in \cite{peng2026signaldependentshotnoisemodeling}. In addition, RAQRs have also attracted growing interest in sensing, such as direction of arrival \cite{11493506} and angle of arrival \cite{11156153}. For applications, owing to their high sensitivity, RAQRs have
demonstrated potential for direct satellite communication
links~\cite{11417150,peng2026celllevelchannelshapingrydberg}
and battery-free communication in ambient Internet of Things
(IoT) networks~\cite{11547166,peng2026rfparadigmrydbergatomic}.

Despite these advances, the sensing capability of RAQRs
remains limited by their instantaneous bandwidth, which is constrained by atomic response times and typically ranges from 1 to 10 MHz \cite{bohaichuk2022origins}. This limitation poses a challenge for high-resolution ranging, which requires a much wider bandwidth. Chirp waveforms offer a natural solution for this challenge. Specifically,  by mixing chirped echoes with a chirped LO inside the atoms, the target delays are converted into low-frequency beat tones, without requiring the atomic receiver to cover the full sweep bandwidth~\cite{watterson2025imaging}. Based on this
principle, a self-heterodyne sensing architecture was
proposed in~\cite{Cuisensing}, where the transmitted chirp
served as both the probing signal and the reference. Extending chirp-based sensing to multicarrier waveforms, affine frequency division multiplexing (AFDM) was investigated for RAQRs \cite{kim2026multichirpafdmrydbergatomic}.

Although a chirped LO enables wideband sensing despite the limited instantaneous bandwidth of RAQRs, its time-varying detuning complicates the atomic response.  As the LO sweeps across a band much wider than the transparency window, its varying detuning shifts the atomic operating point through the AC Stark effect and changes the dressing strength. Consequently, the bias and conversion gain vary within each chirp, resulting in a linear time-varying (LTV) RAQR. The resulting bias variation causes spectral leakage into the
beat band, while gain variation spreads the beat tones and produces nonuniform sensing information. Although the operating point can be stabilized through transmit-power control as in~\cite{Cuisensing}, the dressing strength still decreases along the sweep, leaving the atomic conversion time-varying. Therefore, achieving a linear time-invariant (LTI) atomic response for chirp waveforms remains an open problem.

To address this challenge, we propose a time-invariant RAQR
framework by introducing an auxiliary (AUX) microwave field
on an adjacent Rydberg transition. First, we design a five-level RAQR that exploits the AUX light shift to compensate for variations in the LO light shift. Then, under far detuning, the five-level receiver is approximated by a three-level EIT ladder with a dressed upper state. Building on this architecture,  the LO
and AUX fields can be jointly designed through a Stark-and-gain lock to maintain a constant operating point and dressing strengths. 
The main contributions of this paper are summarized as follows.

\begin{itemize}
\item The equivalent RF-to-optical model of a five-level RAQR is derived for a multicarrier AFDM sensing system. Based on this model, the chirp-induced time-varying bias and conversion gain are characterized, and then the resulting spectral leakage in the range profile and range-Doppler map is analyzed. Furthermore, by analyzing the effects of LO and AUX driving on the atomic operating point, we show that the four-level receiver cannot simultaneously stabilize the operating point and dressing strength, whereas the AUX provides the additional control to overcome this
limitation. 

\item Building on this analysis, we establish the equivalence between the five-level receiver and a dressed three-level EIT ladder under far detuning. This equivalence enables a joint LO-AUX.
Stark-and-gain lock that stabilizes the atomic operating point and conversion gain within each chirp, yielding an effectively LTI atomic response. Consequently, the bias is confined to the direct current (DC) bin and each on-grid beat tone remains in its corresponding range bin,
suppressing receiver-induced spectral leakage and facilitating efficient fast Fourier transform (FFT)-based sensing.

\item Based on the LTI atomic response, the AUX field decouples the operating point from the LO dressing, which remains coupled through the LO amplitude in the four-level receiver. Benefiting from this decoupling, we derive the closed-form expressions for the optimal working point and LO dressing.  Considering the LO and AUX power constraints, we further analyze an upper bound on the conversion gain. Simulation results confirm the accuracy of our derivation and analysis.

\item Under the proposed lock, the impacts of the LO and AUX on the output signal-to-noise ratio (SNR) and the Cram\'er--Rao bounds (CRBs) of delay and Doppler estimation are analyzed. Accounting for power constraints of LO and AUX,  a range-accuracy floor and a delay--Doppler tradeoff are revealed, providing guidelines for LO and AUX design.
Compared with the four-level receiver~\cite{Cuisensing}, the proposed framework improves the SNR while suppressing bias-induced spectral leakage. Consequently, simple FFT-based estimation can approach the CRBs at lower transmit power while maintaining superior sensing accuracy
across the evaluated sweep bandwidths.
\end{itemize}

The remainder of this paper is organized as follows. Section~II presents the system model and RAQR based on a 5-level configuration. Section~III develops LTI atomic conversion via LO-AUX dressing. Section~IV analyzes the sensing performance. Section~V presents simulation results, and Section~VI concludes the paper.

\section{System Model}
\label{sec:system_model}
This section develops a received-signal model that relates the transmitted multicarrier AFDM waveform to the output probe intensity of a RAQR.

\subsection{AFDM Signal}
\label{sec:afdm_signal}

We consider an AFDM waveform with $N$ chirp subcarriers, among which
$N_{\rm s}\le N$ are active. Following \cite{11770275}, the $n$-th entry of
the $m$-th subcarrier is given by
\begin{equation}
\psi_m[n]=\frac{1}{\sqrt N}\,
e^{\mathrm j2\pi\left(c_1n^2+\frac{mn}{N}+c_2m^2\right)},
n,m=0,\cdots,N-1,
\label{eq:idaft_basis}
\end{equation}
where $c_1$ and $c_2$ are the post-chirp and pre-chirp parameters. 
With the sampling bandwidth $B_{\rm s}=N/T$, the continuous-time counterpart
of~\eqref{eq:idaft_basis} is the wrapped chirp, which is expressed as
\begin{equation}
\begin{split}
\Psi_m(t)&=e^{\mathrm j2\pi[c_2m^2+\int_0^tf_m(u)\,{\rm d}u]},\\
f_m(t)&=\big\langle\alpha t+m\Delta f+\tfrac{B_{\rm s}}{2}\big\rangle_{B_{\rm s}}
-\tfrac{B_{\rm s}}{2}.
\end{split}
\label{eq:afdm_subcarrier}
\end{equation}
For $0\le t<T$, where $T$ is the symbol duration, $\Delta f=1/T$ is the
subcarrier spacing, $\alpha=2c_1B_{\rm s}^2$ is the chirp rate, and
$\langle\cdot\rangle_{B_{\rm s}}$ wraps the instantaneous frequency $f_m(t)$
into $[-B_{\rm s}/2,B_{\rm s}/2)$. With the symbols $x[m]$ satisfying
$\sum_m\mathbb E\{|x[m]|^2\}=1$, the baseband waveform is written as
\begin{equation}
s(t)=\sum_{m=0}^{N_{\rm s}-1}x[m]\,\Psi_m(t).
\label{eq:afdm}
\end{equation}

A sensing frame consists of $M$ blocks with the repetition interval
$T_{\rm r}$, each preceded by a chirp-periodic prefix (CPP). Block
$l=1,\cdots,M$ carries the symbols $x_l[m]$ through $s_l(t)$
in~\eqref{eq:afdm}. The transmit waveform is given by
\begin{equation}
x(t)=\sqrt{P_{\rm tx}G_{\rm tx}}\,s(t)\,e^{\mathrm j2\pi f_{\rm c}t},\quad
s(t)=\sum_{l=1}^{M}s_l(t-lT_{\rm r}),
\label{eq:tx}
\end{equation}
where $f_{\rm c}$ is the carrier frequency, $P_{\rm tx}$ is the transmit
power, and $G_{\rm tx}$ is the antenna gain toward the targets.
For target $k$, let $d_{{\rm t},k}$ and $d_{{\rm r},k}$
denote the BS-to-target and target-to-cell distances,
respectively, with total path length
$L_k=d_{{\rm t},k}+d_{{\rm r},k}$. The channel coefficient
is expressed as
\begin{equation}
h_k=
\frac{\lambda_{\rm c}\sqrt{\sigma_k}}
{(4\pi)^{3/2}d_{{\rm t},k}d_{{\rm r},k}}
e^{\mathrm{j}\varphi_k},
\label{eq:hk}
\end{equation}
where $\lambda_{\rm c}=c/f_{\rm c}$ is the carrier
wavelength, $\sigma_k$ is the bistatic radar cross section
(RCS), $\varphi_k$ is the reflection phase, and $c$ is
the speed of light. The received complex echo is written as
\begin{equation}
y_{\rm s}(t)=\sum_{k=1}^{K}h_k\,x(t-\tau_k)\,e^{\mathrm j2\pi\nu_kt},
\label{eq:echo-power}
\end{equation}
where $\tau_k=L_k/c$ is the propagation delay and
$\nu_k=-\frac{f_{\rm c}}{c}L^\prime_k$ is the Doppler shift,
with $L^\prime_k=\frac{\mathrm dL_k}{\mathrm dt}$.\footnote{
The CPP satisfies $T_{\rm cp}>\max_k\tau_k$, ensuring
that echoes received during $0\le t<T$ originate from
the current block or its prefix. The delay is treated
as constant over each frame, and Doppler-induced chirp
time scaling is neglected under
$|L^\prime_k|MT_{\rm r}\ll \frac{c}{|B|}$ and
$\pi N|L^\prime_k|/(2c)\ll1$.}

The corresponding complex electric field at the vapor
cell is expressed as
\begin{equation}
\begin{aligned}
E_{\rm s}(t)
&=\sqrt{\frac{2Z_0}{A_{\rm c}}}\,y_{\rm s}(t)\\
&=\sum_{k=1}^{K}\beta_k\,s(t-\tau_k)\,
e^{\mathrm j2\pi[f_{\rm c}(t-\tau_k)+\nu_kt]},
\end{aligned}
\label{eq:echo}
\end{equation}
where $Z_0\approx377~\Omega$ is the free-space impedance,
$A_{\rm c}=\frac{\lambda_{\rm c}^2}{4\pi}$ is the effective
aperture of the unit-gain reference receiver, and
$\beta_k=\sqrt{\frac{2Z_0P_{\rm tx}G_{\rm tx}}{A_{\rm c}}}\,h_k$
is the complex amplitude of echo $k$.
\subsection{LO Chirp and Composite Field}\label{sec:lo}
The LO chirp received at the vapor cell is the delayed $m=0$ chirp on the same
carrier, which is expressed as
\begin{equation}\label{eq:lo-rx}
\begin{split}
 y_{\rm LO}(t)&=\sqrt{P_{\rm LO}(t)}\;e^{\mathrm j\Theta_{\rm LO}(t)},\\
 \Theta_{\rm LO}(t)&=2\pi\left[f_{\rm c}(t-\tau_{\rm L})
 +\phi_0(t-lT_{\rm r}-\tau_{\rm L})\right].
\end{split}
\end{equation}
For $lT_{\rm r}\le t<lT_{\rm r}+T$, $P_{\rm LO}(t)$ is the received
power, which repeats in every block, and $\tau_{\rm L}$ is the LO delay.
Similar to~\eqref{eq:echo}, the LO field is given by
\begin{equation}\label{eq:lo}
 E_{\rm LO}(t)=\sqrt{\frac{2Z_0}{A_{\rm c}}}\;y_{\rm LO}(t)
 =\lvert E_{\rm LO}(t)\rvert\,e^{\mathrm j\Theta_{\rm LO}(t)}
\end{equation}
where $\lvert E_{\rm LO}(t)\rvert=\sqrt{2Z_0P_{\rm LO}(t)/A_{\rm c}}$ is the electric-field magnitude, and the instantaneous LO frequency is given by
\begin{equation}
 f_{\rm LO}(t)=\frac{1}{2\pi}\frac{{\rm d}\Theta_{\rm LO}(t)}{{\rm d}t}
 =f_{\rm c}+f_0(t-lT_{\rm r}-\tau_{\rm L}).
\end{equation}

Combining~\eqref{eq:lo} with~\eqref{eq:echo}, the superimposed field is
written as
\begin{equation}\label{eq:composite}
 E_{\rm RF}(t)=E_{\rm LO}(t)+E_{\rm s}(t)
 =\widetilde E_{\rm RF}(t)\,e^{\mathrm j\Theta_{\rm LO}(t)},
\end{equation}
where $\widetilde E_{\rm RF}(t)$ is the LO-referenced envelope. Since
$s(t-\tau_k)=s_l(t-lT_{\rm r}-\tau_k)$ within block $l$,
substituting~\eqref{eq:afdm}, \eqref{eq:echo}, and~\eqref{eq:lo}
into~\eqref{eq:composite}, we have
\begin{equation}\label{eq:envelope}
\begin{aligned}
 \widetilde E_{\rm RF}(t)&=\lvert E_{\rm LO}(t)\rvert
 +\sum_{k=1}^{K}\sum_{m=0}^{N_{\rm s}-1}b_{k,m,l}\,
 e^{\mathrm j2\pi\Delta\phi_{k,m}(t-lT_{\rm r})},\\
 b_{k,m,l}&=\beta_k\,x_l[m]\,
 e^{\mathrm j2\pi\left[c_2m^2-f_{\rm c}(\tau_k-\tau_{\rm L})
 +\nu_klT_{\rm r}\right]},\\
 \Delta\phi_{k,m}(t)&=f_{k,m}t+\psi_{k,m}.
\end{aligned}
\end{equation}
Here, $b_{k,m,l}$ collects the time-independent complex amplitude factors of
block $l$, and $\Delta\phi_{k,m}(t)$ is the LO-referenced phase in cycles. The
matched sweep rates cancel the quadratic phase terms, leaving a constant beat
frequency and phase offset, which are given by
\begin{equation}\label{eq:beat}
\begin{aligned}
 f_{k,m}&\triangleq m\Delta f-\alpha(\tau_k-\tau_{\rm L})+\nu_k,\\
 \psi_{k,m}&\triangleq\frac{\alpha}{2}\big(\tau_k^2-\tau_{\rm L}^2\big)
 -m\Delta f\,\tau_k.
\end{aligned}
\end{equation}


\subsection{Five-Level Atomic Response}\label{sec:atom}
We consider the 5-level configuration with \(\{|1\rangle,|2\rangle,|3\rangle,|4\rangle,|5\rangle\}\),
where \(|1\rangle\) denotes the ground state, \(|2\rangle\) represents the
excited state, and \(|3\rangle\), \(|4\rangle\), and \(|5\rangle\) are Rydberg
states. The probe and coupling lasers drive
\(|1\rangle\rightarrow|2\rangle\) and \(|2\rangle\rightarrow|3\rangle\)
with the constant Rabi frequencies \(\Omega_{\rm p}\) and \(\Omega_{\rm c}\),
and both lasers are tuned exactly on resonance, namely
\(\Delta_{\rm p}=\Delta_{\rm c}=0\). The superimposed signal of LO and echo drives \(|3\rangle\rightarrow|4\rangle\) with the complex Rabi
frequency \(\Omega_{\rm RF}(t)=\Omega_{\rm LO}(t)+\Omega_{\rm s}(t)\), where
\(\Omega_{\rm LO}(t)=\frac{\mu_{34}\lvert E_{\rm LO}(t)\rvert}{\hbar}\) and
\(\Omega_{\rm s}(t)=\frac{\mu_{34}}{\hbar}\sum_{k=1}^{K}\sum_{m=0}^{N_{\rm s}-1}
b_{k,m}\,e^{\mathrm j2\pi\Delta\phi_{k,m}(t)}\). In addition, an AUX field
\(E_{\rm A}(t)\cos\theta_{\rm A}(t)\), with the amplitude \(E_{\rm A}(t)\) and
the instantaneous angular frequency
\(\omega_{\rm A}(t)=\theta_{\rm A}^{\prime}(t)\), drives
\(|3\rangle\rightarrow|5\rangle\) with the Rabi frequency
\(\Omega_{\rm A}(t)=\frac{\mu_{35}E_{\rm A}(t)}{\hbar}\). Here, \(\mu_{ij}\)
denotes the dipole moment between \(|i\rangle\) and \(|j\rangle\), and
\(\hbar\) represents the reduced Planck constant. Furthermore, we assume that \(\Omega_{\rm p}\), \(\Omega_{\rm c}\), \(\Omega_{\rm LO}(t)\), and
\(\Omega_{\rm A}(t)\) are real-valued and nonnegative, as the constant
phases of all fields are absorbed into the atomic basis.

By applying the rotating-wave approximation (RWA), the time-dependent
Hamiltonian is given by
\begin{equation}\label{eq:H_full}
 \mathbf H(t)=\frac{\hbar}{2}
 \begin{bmatrix}
 0 & \Omega_{\rm p} & 0 & 0 & 0\\
 \Omega_{\rm p} & 0 & \Omega_{\rm c} & 0 & 0\\
 0 & \Omega_{\rm c} & 2\varepsilon_3(t) & \Omega_{\rm RF}(t) & \Omega_{\rm A}(t)\\
 0 & 0 & \Omega_{\rm RF}^{*}(t) & 2\varepsilon_4(t) & 0\\
 0 & 0 & \Omega_{\rm A}(t) & 0 & 2\varepsilon_5(t)
 \end{bmatrix},
\end{equation}
where $\varepsilon_3(t)=\delta_3(t)$,
$\varepsilon_4(t)=\Delta_{\rm LO}(t)+\delta_4(t)$, and
$\varepsilon_5(t)=-\Delta_{\rm A}(t)+\delta_5(t)$ are the
rotating-frame energies of $|3\rangle$, $|4\rangle$, and
$|5\rangle$, respectively, expressed in angular-frequency units.
The LO and AUX detunings are defined as
$\Delta_{\rm LO}(t)=\omega_{\rm LO}(t)-\omega_{34}$ and
$\Delta_{\rm A}(t)=\omega_{\rm A}(t)-\omega_{53}$, where
$\omega_{\rm LO}(t)=\Theta_{\rm LO}'(t)$ is the instantaneous
LO angular frequency, and $\omega_{34}$ and $\omega_{53}$
are the transition angular frequencies of
$|3\rangle\rightarrow|4\rangle$ and
$|5\rangle\rightarrow|3\rangle$, respectively.
The term $\delta_a(t)$, $a\in\{3,4,5\}$, denotes the AC Stark
shift of $|a\rangle$ induced by off-resonant dipole couplings
to the LO and AUX fields. Thus, the time dependence of $\mathbf H(t)$ arises from the
chirped LO and echoes through $\Omega_{\rm RF}(t)$ and
$\Delta_{\rm LO}(t)$, AUX control through $\Omega_{\rm A}(t)$
and $\Delta_{\rm A}(t)$, and the induced AC Stark shifts
$\delta_a(t)$.

Let $\boldsymbol{\rho}(t)\in\mathbb C^{5\times5}$ denote the
atomic density matrix, with entries $\rho_{ij}(t)$. Its
evolution is governed by the Lindblad master equation
\begin{equation}
\frac{\partial \boldsymbol{\rho}(t)}{\partial t}
=-\frac{\mathrm{j}}{\hbar}
[\mathbf H(t),\boldsymbol{\rho}(t)]
+\mathcal L[\boldsymbol{\rho}(t)],
\label{eq:mastereq}
\end{equation}
where
$\mathcal L[\boldsymbol{\rho}]
=-\frac{1}{2}\{\boldsymbol{\Gamma},\boldsymbol{\rho}\}
+\boldsymbol{\Lambda}(\boldsymbol{\rho})$
describes population decay and repopulation, with \(\boldsymbol{\Gamma}
=\operatorname{diag}\{0,\gamma_2,\gamma_3,\gamma_4,\gamma_5\}\) and \(\boldsymbol{\Lambda}(\boldsymbol{\rho}) = \operatorname{diag}\big\{
\gamma_2\rho_{22}+\gamma_4\rho_{44}+\gamma_5\rho_{55},\gamma_3\rho_{33},0,0,0\big\}\). $\gamma_i$, $i=2,3,4,5$, denotes the population decay
rate of state $|i\rangle$. For the considered cold atoms, Doppler and transit-time broadening are assumed negligible, and relaxation is modeled solely through
these decay processes \cite{Gongtcom,Cuisensing}.

Following~\cite{Cuisensing}, we adopt a quasi-static response model
and neglect Rydberg-state decay by setting
$\gamma_3=\gamma_4=\gamma_5=0$.\footnote{The rotating-frame drive
envelopes and effective detunings, including the echo--LO beat
dynamics, are assumed to vary slowly on the atomic relaxation
timescale after initial transients have decayed.} The imaginary of the steady-state response is expressed as
\begin{equation}
\Im\{\rho_{21}(t)\}
=-\frac{B_1Y^2(t)}{D(t)},
\label{eq:rho21}
\end{equation}
where
\begin{equation}
\begin{aligned}
Y(t)=&|\Omega_{\rm RF}(t)|^2\varepsilon_5(t)
+\Omega_{\rm A}^2(t)\varepsilon_4(t)-4\varepsilon_3(t)\varepsilon_4(t)\varepsilon_5(t),
\end{aligned}
\label{eq:Y}
\end{equation}
\begin{equation}
\begin{aligned}
D(t)=&C_1Y^2(t)
+C_2\Big[
|\Omega_{\rm RF}(t)|^2\varepsilon_5^2(t)
+\Omega_{\rm A}^2(t)\varepsilon_4^2(t)
\Big]\\
+&C_3\varepsilon_4^2(t)\varepsilon_5^2(t)+\frac{\Omega_{\rm p}^4}{4}
\Big\{
\big[|\Omega_{\rm RF}(t)|^2+\Omega_{\rm A}^2(t)\big]^2\\
-&8\varepsilon_3(t)
\big[
|\Omega_{\rm RF}(t)|^2\varepsilon_4(t)
+\Omega_{\rm A}^2(t)\varepsilon_5(t)
\big]\\
+&16\varepsilon_3^2(t)
\big[\varepsilon_4^2(t)+\varepsilon_5^2(t)\big]
\Big\},
\end{aligned}
\label{eq:D}
\end{equation}
and
$B_1=\gamma_2\Omega_{\rm p}$,
$C_1=2\Omega_{\rm p}^2+\gamma_2^2$,
$C_2=2\Omega_{\rm p}^2(\Omega_{\rm c}^2+\Omega_{\rm p}^2)$,
and $C_3=4(\Omega_{\rm c}^2+\Omega_{\rm p}^2)^2$.
Removing $|5\rangle$ and setting $\varepsilon_3(t)=0$, we obtain the
result based on the four-level RAQR~\cite{Cuisensing}.

Evaluating the susceptibility at the input probe Rabi frequency,
the Beer--Lambert model gives the output probe power as
\begin{equation}
P_{\rm out}(t)= P_{\rm in}
\exp\!\left[
L_{\rm cell}k_{\rm p}D_{\Omega}
\Im\{\rho_{21}(t)\}
\right],
\label{outprobe}
\end{equation}
where $P_{\rm in}$ is the input probe power,
$L_{\rm cell}$ is the atomic-medium length, and
$k_{\rm p}=\frac{2\pi}{\lambda_{\rm p}}$ is the probe wavenumber.
The coefficient
$D_{\Omega}=\frac{2N_0\mu_{12}^2}{\epsilon_0\hbar\Omega_{\rm p}}$
relates the probe coherence to the optical susceptibility,
with $N_0$ denoting the atomic number density.


\section{LTI Receiver Design via LO--AUX Dressing}\label{sec:design}
In this section, we first derive the equivalent baseband signal and then
characterize the impacts of the LO and AUX fields on it. Based on these
results, we design a Stark-and-gain lock that renders the LTI atomic response.
 
\subsection{Equivalent Baseband Signal}\label{sec:baseband}
Based on the output probe power intensity, we derive the equivalent baseband signal. First, since the echoes are typically weaker than the LO, namely
\(\lvert\Omega_{\rm s}(t)\rvert\ll\Omega_{\rm LO}(t)\), we have 
\begin{equation}\label{eq:OmegaRF2}
 \lvert\Omega_{\rm RF}(t)\rvert^2\approx\Omega_{\rm LO}^2(t)
 +2\Omega_{\rm LO}(t)\Re\{\Omega_{\rm s}(t)\}.
\end{equation}
Then, by using the Taylor expansion at \(\Omega_{\rm LO}(t)\), the output probe intensity is approximated as \cite{Cuisensing}
\begin{equation}\label{appprobe}
    P_{\rm out}(t)\approx\Pi(t)+g(t)\,\Re\{\Omega_{\rm s}(t)\}
\end{equation}
where \( \Pi(t)\) is the time-varying bias, given by
\begin{equation}\label{eq:bias}
 \Pi(t)=P_{\rm in}\exp\!\Big[L_{\rm cell}k_{\rm p}D_{\Omega}
 \Im\{\rho_{21}(t)\}\big|_{\Omega_{\rm RF}(t)=\Omega_{\rm LO}(t)}\Big].
\end{equation}
\(g(t)=\frac{\partial\Pi(t)}{\partial\Omega_{\rm LO}(t)}\) is the conversion gain, written as
\begin{equation}\label{eq:gain}
\begin{aligned}
 g(t)=&-2\Omega_{\rm LO}(t)\,\Pi(t)\,L_{\rm cell}k_{\rm p}D_{\Omega}B_1\\
 \times&\Bigg[\frac{Y(t)\Big(2\varepsilon_5(t)D(t)
 -Y(t)\frac{\partial D(t)}{\partial\lvert\Omega_{\rm RF}(t)\rvert^2}\Big)}
 {D^2(t)}\Bigg]_{\Omega_{\rm RF}(t)=\Omega_{\rm LO}(t)},
\end{aligned}
\end{equation}
where
\begin{equation}\label{eq:dD}
\begin{aligned}
 \frac{\partial D(t)}{\partial\lvert\Omega_{\rm RF}(t)\rvert^2}
 =&2C_1Y(t)\varepsilon_5(t)+C_2\varepsilon_5^2(t)\\
 +&\frac{\Omega_{\rm p}^4}{2}\big[\lvert\Omega_{\rm RF}(t)\rvert^2
 +\Omega_{\rm A}^2(t)-4\varepsilon_3(t)\varepsilon_4(t)\big].
\end{aligned}
\end{equation}
Therefore, \(\Pi(t)\) and \(g(t)\)
are determined by the operating point, namely \(\Omega_{\rm LO}(t)\),
\(\Omega_{\rm A}(t)\), \(\varepsilon_3(t)\), \(\varepsilon_4(t)\), and
\(\varepsilon_5(t)\), rather than the echoes.
 
Substituting \(\Omega_{\rm s}(t)\) with
\(\Delta\phi_{k,m}(t-lT_{\rm r})=f_{k,m}(t-lT_{\rm r})+\psi_{k,m}\) into~\eqref{appprobe}, we have 
\begin{equation}\label{eq:baseband_tones}
P_{\rm out}(t)\approx\Pi(t)+\Re\bigg\{g(t)\sum_{k=1}^{K}\sum_{m=0}^{N_{\rm s}-1}
\Omega_{k,m,l}\,e^{\mathrm j2\pi f_{k,m}(t-lT_{\rm r})}\bigg\},
\end{equation}
where \(\Omega_{k,m,l}=\frac{\mu_{34}}{\hbar}b_{k,m,l}e^{\mathrm j2\pi\psi_{k,m}}\)
is the complex Rabi amplitude of the beat tone of target \(k\) on subcarrier
\(m\). Therefore, within one chirp, the bias and the amplitude of every beat
tone vary with time through \(\Pi(t)\) and \(g(t)\), whereas an LTI receiver
requires both to be constant.

The photodetector converts the probe output into a photocurrent with the
responsivity \(\mathcal R_{\rm ph}\), which flows through a unit resistance and is amplified with the power gain \(G_{\rm ph}\).
After photodetection, a low-pass filter (LPF) with cutoff frequency
\(f_{\rm LP}>\max\limits_{k,m}f_{k,m}\) is applied to suppress the out-of-band noise, and then the output is sampled with the sampling rate \(\frac{N_{\rm FFT}}{T}\geq2f_{\rm LP}\). Then, the \(i\)-th sample of chirp \(l\) is expressed as
\begin{equation}\label{eq:samples}
\begin{split}
     V_l[i] & = \sqrt{G_{\rm ph}}\mathcal{R}_{\rm ph} P_l[i] + w_l[i],\\
     P_l[i]&\approx\Pi[i]+\Re\bigg\{g[i]\sum_{k=1}^{K}\sum_{m=0}^{N_{\rm s}-1}
 \Omega_{k,m,l}\,e^{\mathrm j2\pi n_{k,m}\frac{i}{N_{\rm FFT}}}\bigg\},
\end{split}
\end{equation}
where \(i=0,\cdots,N_{\rm FFT}-1\), \(l=0,\cdots,M-1\), \(\Omega_{k,m,l}\) is the
beat amplitude in chirp \(l\), and
\(n_{k,m}=f_{k,m}T\in\big(0,\frac{N_{\rm FFT}}{2}\big)\) is the beat bin of
target \(k\) on subcarrier \(m\). $w_l[i]$ denotes the noise term, which is characterized in the following section.
\(\Pi[i]=\Pi\big(\frac{iT}{N_{\rm FFT}}\big)\) and
\(g[i]=g\big(\frac{iT}{N_{\rm FFT}}\big)\) are identical across chirps, since
the LO and AUX programs repeat in every chirp. The range profile and the
range-Doppler map can be obtained by the following lemma.

\begin{lemma}[Range profile and range-Doppler map]\label{lem:fft}
Let \(\hat P_l[n]\) be the \(N_{\rm FFT}\)-point FFT of \(P_l[i]\) over \(i\),
normalized by \(\frac{1}{N_{\rm FFT}}\), namely the range profile of chirp
\(l\), and let \(\hat P[n,n_{\rm D}]\) be the \(M\)-point FFT of
\(\hat P_l[n]\) over \(l\), normalized by \(\frac{1}{M}\), namely the
range-Doppler map, where \(n\) and \(n_{\rm D}\) are the range and Doppler bins,
respectively. Then, we have
\begin{equation}\label{eq:leak_fft}
\begin{aligned}
 \hat P_l[n]=&\hat\Pi[n]+\frac{1}{2}\sum_{k=1}^{K}\sum_{m=0}^{N_{\rm s}-1}
 \Big[\Omega_{k,m,l}\,\hat g(n-n_{k,m})\\
 +&\Omega_{k,m,l}^{*}\,\hat g(n+n_{k,m})\Big],
\end{aligned}
\end{equation}
\begin{equation}\label{eq:leak_2d}
\begin{aligned}
 \hat P[n,n_{\rm D}]=&\hat\Pi[n]\,\delta[n_{\rm D}]
 +\frac{1}{2}\sum_{k=1}^{K}\sum_{m=0}^{N_{\rm s}-1}
 \Big[\check\Omega_{k,m}[n_{\rm D}]\\
\times &\hat g(n-n_{k,m})+\check\Omega_{k,m}^{*}[-n_{\rm D}]\,
 \hat g(n+n_{k,m})\Big],
\end{aligned}
\end{equation}
where \(\hat\Pi[n]\) is the FFT of \(\Pi[i]\) normalized as above,
\(\hat g(\zeta)=\frac{1}{N_{\rm FFT}}\sum_{i=0}^{N_{\rm FFT}-1}g[i]\,
e^{-\mathrm j2\pi\frac{\zeta i}{N_{\rm FFT}}}\) is the spectrum of \(g[i]\) at
the real-valued bin \(\zeta\), which coincides with its FFT at integer
\(\zeta\), \(\check\Omega_{k,m}[n_{\rm D}]\) is the FFT of \(\Omega_{k,m,l}\)
over \(l\) normalized as above, the Doppler bins are taken modulo \(M\), and
\(\delta[\cdot]\) is the Kronecker delta.
\end{lemma}

\begin{IEEEproof}
Since \(g[i]\) is real, we have
\(\Re\{g[i]\,\Omega_{k,m,l}e^{\mathrm j2\pi\frac{n_{k,m}i}{N_{\rm FFT}}}\}
=\frac{g[i]}{2}\big(\Omega_{k,m,l}\,e^{\mathrm j2\pi\frac{n_{k,m}i}{N_{\rm FFT}}}
+\Omega_{k,m,l}^{*}\,e^{-\mathrm j2\pi\frac{n_{k,m}i}{N_{\rm FFT}}}\big)\). For
any \(n_{k,m}\), the normalized FFT maps
\(g[i]e^{\pm\mathrm j2\pi\frac{n_{k,m}i}{N_{\rm FFT}}}\) to
\(\hat g(n\mp n_{k,m})\). Summing these results and adding the bias spectrum $\hat\Pi[n]$, we obtain \eqref{eq:leak_fft}. Then, since \(\hat\Pi[n]\) and \(\hat g(\zeta)\) are independent of \(l\), the FFT over \(l\) acts only on the constant factor \(1\),
\(\Omega_{k,m,l}\), and \(\Omega_{k,m,l}^{*}\), and thus we have
\(\delta[n_{\rm D}]\), \(\check\Omega_{k,m}[n_{\rm D}]\), and
\(\check\Omega_{k,m}^{*}[-n_{\rm D}]\). Therefore, we complete the proof of \eqref{eq:leak_2d}.
\end{IEEEproof}

\begin{remark}[Spectral leakage of a time-varying receiver]\label{rem:leakage}
For constant bias and gain, Lemma~\ref{lem:fft} shows that the bias
contributes only a removable DC component, while each beat tone appears as the Dirichlet kernel
\(\mathcal W(n-n_{k,m})\) of the observation window centered at its
positive-frequency beat bin \(n_{k,m}\), where \(\mathcal W(\zeta)=\frac{1}{N_{\rm FFT}}
\sum_{i=0}^{N_{\rm FFT}-1}e^{-\mathrm j2\pi\frac{\zeta i}{N_{\rm FFT}}}\) satisfies
\(\mathcal W(n)=\delta[n]\) at integer bins. In contrast, within-chirp bias variations introduce
a non-DC background, while gain variations modulate the beat tones
and spread their energy across range bins. Because these variations
repeat identically across chirps, the bias background remains at
zero Doppler, and gain modulation introduces no additional Doppler
spreading. The resulting non-DC leakage within the beat band survives
low-pass filtering and DC blocking, potentially masking weak targets
and degrading sensing accuracy.
\end{remark}
 
Remark~\ref{rem:leakage} shows that the leakage originates from the time
variation of \(\Pi(t)\) and \(g(t)\) within one chirp. In~\cite{Cuisensing}, the transmit-power trajectory is optimized to improve SNR across the chirp in a self-referenced receiver. This approach can increase PAPR, while the time-varying atomic
response requires accurate bias subtraction and gain-aware estimation. To tackle these challenges, we first analyze how LO and AUX driving affect the bias and gain, and then jointly design their control profiles to keep both constant within each chirp. This allows DC blocking to remove the bias and suppresses gain-induced leakage, simplifying FFT-based sensing.
 
\subsection{Impacts of LO and AUX on atomic response}\label{sec:impacts}

To reveal the impacts of the  LO and AUX on the bias and atomic gain, we first define the dressing ratios as
\begin{equation}\label{eq:ratios}
 a_{\rm LO}(t)=\frac{\Omega_{\rm LO}(t)}{2\lvert\varepsilon_4(t)\rvert},\qquad
 a_{\rm A}(t)=\frac{\Omega_{\rm A}(t)}{2\lvert\varepsilon_5(t)\rvert}.
\end{equation}
Then, the effective detuning of \(|3\rangle\) is denoted as
\begin{equation}\label{eq:Deff}
 \Delta_{\rm eff}(t)=\varepsilon_3(t)-a_{\rm LO}^2(t)\varepsilon_4(t)
 -a_{\rm A}^2(t)\varepsilon_5(t),
\end{equation}
where \(-a_{\rm LO}^2(t)\varepsilon_4(t)=-\frac{\Omega_{\rm LO}^2(t)}{4\varepsilon_4(t)}\)
and \(-a_{\rm A}^2(t)\varepsilon_5(t)=-\frac{\Omega_{\rm A}^2(t)}{4\varepsilon_5(t)}\)
are the light shifts of \(|3\rangle\) induced by the LO through \(|4\rangle\)
and by the AUX through \(|5\rangle\), respectively. Hence,
\(\Delta_{\rm eff}(t)\) is the detuning of the probe and coupling lasers from
the dressed \(|3\rangle\), which includes the AC Stark shift
\(\varepsilon_3(t)\). \(\Delta_{\rm eff}(t)=0\) means that the probe and coupling lasers are resonant
with the dressed \(|3\rangle\).
 
At the LO operating point, substituting~\eqref{eq:Deff} into~\eqref{eq:Y}, we have \(Y(t)=-4\varepsilon_4(t)\varepsilon_5(t)\Delta_{\rm eff}(t)\). Similarly, substituting \(\Omega_{\rm LO}^2(t)=4a_{\rm LO}^2(t)\varepsilon_4^2(t)\), \(\Omega_{\rm A}^2(t)=4a_{\rm A}^2(t)\varepsilon_5^2(t)\), and \(\varepsilon_3(t)=\Delta_{\rm eff}(t)+a_{\rm LO}^2(t)\varepsilon_4(t) +a_{\rm A}^2(t)\varepsilon_5(t)\) into~\eqref{eq:D}, the
response~\eqref{eq:rho21} can be rewritten as
\begin{equation}\label{eq:rho21_exact}
 \Im\{\rho_{21}(t)\}=-\frac{B_1\Delta_{\rm eff}^2(t)}
 {C_1\Delta_{\rm eff}^2(t)+Q(t)+E(t)},
\end{equation}
where
\begin{equation}\label{eq:Q}
 Q(t)=\frac{1}{4}\Big[\Omega_{\rm c}^2
 +\big(1+a_{\rm LO}^2(t)+a_{\rm A}^2(t)\big)\Omega_{\rm p}^2\Big]^2,
\end{equation}
\begin{equation}\label{eq:E}
\begin{aligned}
 E(t)=&\frac{\Omega_{\rm p}^4}{4}\Bigg[
 2\Delta_{\rm eff}(t)\left(\frac{a_{\rm LO}^2(t)}{\varepsilon_4(t)}
 +\frac{a_{\rm A}^2(t)}{\varepsilon_5(t)}\right)\\
 +&\Delta_{\rm eff}^2(t)\left(\frac{1}{\varepsilon_4^2(t)}
 +\frac{1}{\varepsilon_5^2(t)}\right)\Bigg].
\end{aligned}
\end{equation}
Here, \(Q(t)\) depends only on the dressing ratios, whereas \(E(t)\) collects
all the terms that depend on \(\varepsilon_4(t)\) and \(\varepsilon_5(t)\)
explicitly.

Since \eqref{eq:rho21_exact} is analytically intractable, we simplify it by choosing a suitable operating regime. In chirp sensing, the LO detuning $\varepsilon_4(t)$ sweeps across a band much wider than $\Omega_{\rm p}$. The sweep is placed entirely on one side of the $|3\rangle\rightarrow|4\rangle$ transition, with $\min_t|\varepsilon_4(t)|\gg\Omega_{\rm p}$. Furthermore, we assume that
the AUX field is operated far from the $|3\rangle\rightarrow|5\rangle$ resonance to simplify the analysis and the locking design. Therefore, we
consider a far-detuning regime \( \lvert\varepsilon_4(t)\rvert,\ \lvert\varepsilon_5(t)\rvert\gg\Omega_{\rm p}\).

\begin{figure}
    \centering
    \includegraphics[width=1\linewidth]{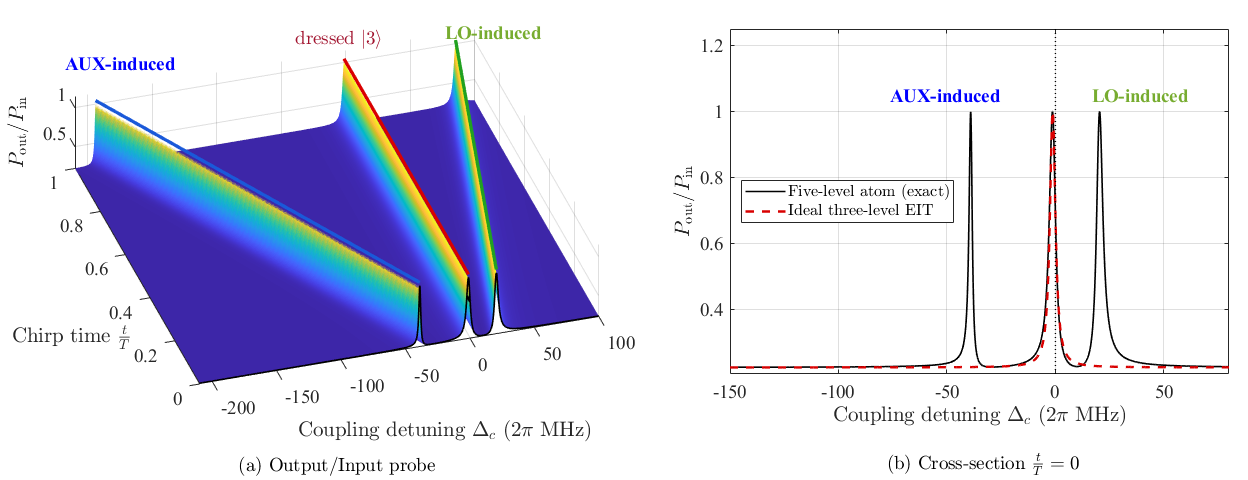}
    \caption{5-level enabled equivalent EIT.}
    \label{Three-level}
\end{figure}

Under far-detuned LO and AUX driving, the dressed modes dominated
by $|4\rangle$ and $|5\rangle$ remain far from the fixed laser
resonance $\Delta_{\rm p}=\Delta_{\rm c}=0$.
Their weak excitation allows the near-resonant response to be
approximated by a three-level EIT ladder
$|1\rangle$--$|2\rangle$--dressed $|3\rangle$, as shown in Fig. \ref{Three-level}. 
This equivalence is formalized in the following proposition.

\begin{proposition}[Three-level equivalence]\label{prop:three}
Let \(\varepsilon_{\min}(t)=\min\{\lvert\varepsilon_4(t)\rvert,
\lvert\varepsilon_5(t)\rvert\}>0\) and
\(\eta(t)=\frac{E(t)}{C_1\Delta_{\rm eff}^2(t)+Q(t)}\). Then, the
response~\eqref{eq:rho21_exact} can be rewritten as
\begin{equation}\label{eq:rho21_eta}
 \Im\{\rho_{21}(t)\}=-\frac{B_1\Delta_{\rm eff}^2(t)}
 {\big[C_1\Delta_{\rm eff}^2(t)+Q(t)\big]\big[1+\eta(t)\big]},
\end{equation}
where
\begin{equation}\label{eq:E_bound}
 \lvert\eta(t)\rvert\leq\frac{\Omega_{\rm p}}{2\sqrt2\,\varepsilon_{\min}(t)}
 +\frac{\Omega_{\rm p}^2}{4\varepsilon_{\min}^2(t)}.
\end{equation}
Under far detuning, the leading-order response obtained by setting
$\eta(t)=0$ coincides with that of the effective three-level ladder
$|1\rangle$--$|2\rangle$--dressed $|3\rangle$.
The dressed upper state has a bare-$|3\rangle$ weight of
$\frac{1}{1+a_{\rm LO}^2(t)+a_{\rm A}^2(t)}$ and the corresponding effective two-photon detuning and coupling Rabi
frequency are given by
\begin{equation}
\begin{aligned}
\tilde\Delta(t)
&=\frac{\Delta_{\rm eff}(t)}
{1+a_{\rm LO}^2(t)+a_{\rm A}^2(t)},\\
\tilde\Omega_{\rm c}(t)
&=\frac{\Omega_{\rm c}}
{\sqrt{1+a_{\rm LO}^2(t)+a_{\rm A}^2(t)}}.
\end{aligned}
\label{eq:dressed_par}
\end{equation}
\end{proposition}

\begin{IEEEproof}
See Appendix \ref{Proofthree}.
\end{IEEEproof}

Based on Proposition \ref{prop:three}, the response under the far-detuning case is given by
\begin{equation}\label{eq:rho21_dress}
 \Im\{\rho_{21}(t)\}\approx-\frac{B_1\Delta_{\rm eff}^2(t)}
 {C_1\Delta_{\rm eff}^2(t)+Q(t)}.
\end{equation}
Its absorption vanishes at
\(\Delta_{\rm eff}(t)=0\) and approaches \(\frac{B_1}{C_1}\) for
\(\lvert\Delta_{\rm eff}(t)\rvert\gg\sqrt{\frac{Q(t)}{C_1}}\),  where
\(\sqrt{\frac{Q(t)}{C_1}}\) is the half-width of the transparency window. Dressing broadens this window through $Q(t)$,
while $\Delta_{\rm eff}(t)$ sets the operating point of the
fixed lasers. Then, the atomic gain \(g(t)\) is approximated as
\begin{equation}\label{gtfar}
    g(t)=\Pi(t)L_{\rm cell}k_{\rm p}D_{\Omega}
\frac{\partial\Im\{\rho_{21}(t)\}}
{\partial\Omega_{\rm LO}(t)}.
\end{equation}
Using~\eqref{eq:rho21_dress} and \(\frac{\partial\Delta_{\rm eff}(t)}
{\partial\Omega_{\rm LO}(t)}
=-a_{\rm LO}(t)\operatorname{sgn}\{\varepsilon_4(t)\}\), \eqref{gtfar} is rewritten as
\begin{equation}
\begin{aligned}
g(t)\approx&
\Pi(t)L_{\rm cell}k_{\rm p}D_{\Omega}B_1
a_{\rm LO}(t)\operatorname{sgn}\{\varepsilon_4(t)\}\\
\times&
\frac{2\Delta_{\rm eff}(t)Q(t)}
{\big[C_1\Delta_{\rm eff}^2(t)+Q(t)\big]^2}.
\end{aligned}
\label{eq:gain_dress}
\end{equation}

Physically, the gain is governed by the motion of the operating point across the transparency window, mainly through $\Delta_{\rm eff}(t)$, while $a_{\rm LO}(t)$ and $a_{\rm A}(t)$ determine the window width through $Q(t)$, with $a_{\rm LO}(t)$ additionally scaling the gain. The residual contributions from the LO-induced variation of $Q(t)$ and from $\eta(t)$ are negligible under the far-detuning case.

The four-level receiver in~\cite{Cuisensing} is recovered from~\eqref{eq:rho21_exact} by setting $a_{\rm A}(t)=0$ and $|\varepsilon_5(t)|\to\infty$, yielding the same equivalent three-level ladder with $\Delta_{\rm eff}(t)=\varepsilon_3(t)-a_{\rm LO}^2(t)\varepsilon_4(t)$. However, the four-level architecture relies only on $\Omega_{\rm LO}(t)$ to control both $\Delta_{\rm eff}(t)$ and $a_{\rm LO}(t)$. Since $\varepsilon_4(t)$ varies along the chirp, these two quantities cannot generally be kept constant simultaneously, resulting in a time-varying $\Pi(t)$ and $g(t)$. In contrast, the five-level receiver overcomes this limitation by introducing the AUX amplitude and frequency as additional controls. Specifically, by designing the LO and AUX light shifts with opposite signs, the AUX can compensate for the variations of LO light shift and $\varepsilon_3(t)$. Consequently, $\Omega_{\rm LO}(t)$, $\Omega_{\rm A}(t)$, and $\omega_{\rm A}(t)$ can jointly keep $\Delta_{\rm eff}(t)$, $a_{\rm LO}(t)$, and $a_{\rm A}(t)$ constant, making $Q(t)$ and the equivalent EIT ladder time invariant along the chirp.  This motivates the following locking design, which removes the leakage identified in Remark~\ref{rem:leakage}.

\subsection{Stark-and-Gain Lock Design}\label{sec:lock}
The locking design requires the following conditions, which are given by
\begin{equation}\label{eq:lockcond}
\Delta_{\rm eff}(t)=\Delta_0,\quad
a_{\rm LO}(t)=a_{{\rm LO},0},\quad
a_{\rm A}(t)=a_{{\rm A},0},
\end{equation}
where $\Delta_0$, $a_{{\rm LO},0}$, and $a_{{\rm A},0}$
are prescribed constants. The first condition provides Stark locking
by fixing the effective detuning, while the remaining two provide
gain locking by maintaining constant LO and AUX dressing strengths.
The LO chirp is further confined to one side of the
$|3\rangle\leftrightarrow|4\rangle$ transition, ensuring a constant
$\operatorname{sgn}\{\varepsilon_4(t)\}$ and
$\min_t|\varepsilon_4(t)|\gg\Omega_{\rm p}$.
 
Substituting~\eqref{eq:ratios} and~\eqref{eq:Deff}
into~\eqref{eq:lockcond}, the LO and AUX control laws are expressed as
\begin{equation}\label{eq:lock}
\begin{aligned}
 \Omega_{\rm LO}(t)&=2a_{{\rm LO},0}\lvert\varepsilon_4(t)\rvert,\\
 \varepsilon_5(t)&=\frac{\varepsilon_3(t)-a_{{\rm LO},0}^2\varepsilon_4(t)-\Delta_0}
 {a_{{\rm A},0}^2},\\
 \Omega_{\rm A}(t)&=2a_{{\rm A},0}\lvert\varepsilon_5(t)\rvert,\\
 \omega_{\rm A}(t)&=\omega_{53}+\delta_5(t)-\varepsilon_5(t).
\end{aligned}
\end{equation}
Accordingly, the LO
amplitude follows the chirp detuning, the AUX frequency places \(|5\rangle\)
such that the AUX light shift cancels the variations of the LO light shift and
of the AC Stark shift, and the AUX amplitude keeps the dressing by \(|5\rangle\)
fixed. Since \(\lvert\varepsilon_5(t)\rvert\approx
\frac{a_{{\rm LO},0}^2}{a_{{\rm A},0}^2}\lvert\varepsilon_4(t)\rvert\), the
condition \(\lvert\varepsilon_5(t)\rvert\gg\Omega_{\rm p}\) also holds for
\(a_{{\rm A},0}<a_{{\rm LO},0}\). The resulting signal model is established in the following theorem.
 
\begin{theorem}[Stark-and-gain lock]\label{thm:lock}
Under the far-detuning case, if the LO and AUX follow~\eqref{eq:lock} with
\(\lvert\Delta_0\rvert\leq\sqrt{\frac{Q_0}{C_1}}\),
then the bias and the gain are time invariant, namely
\(\Pi(t)\approx\Pi_0\) and \(g(t)\approx g_0\), where
\begin{equation}\label{eq:Pi0g0}
\begin{aligned}
 \Pi_0&=P_{\rm in}\exp\!\bigg[-L_{\rm cell}k_{\rm p}D_{\Omega}
 \frac{B_1\Delta_0^2}{C_1\Delta_0^2+Q_0}\bigg],\\
 g_0&=\Pi_0L_{\rm cell}k_{\rm p}D_{\Omega}B_1\,a_{{\rm LO},0}
 \operatorname{sgn}\{\varepsilon_4\}\frac{2\Delta_0Q_0}{\big(C_1\Delta_0^2+Q_0\big)^2},
\end{aligned}
\end{equation}
with
\begin{equation}\label{eq:Q0}
 Q_0=\frac{1}{4}\Big[\Omega_{\rm c}^2
 +\big(1+a_{{\rm LO},0}^2+a_{{\rm A},0}^2\big)\Omega_{\rm p}^2\Big]^2.
\end{equation}
\end{theorem}
 
\begin{IEEEproof}
Substituting~\eqref{eq:lock} into~\eqref{eq:ratios} and~\eqref{eq:Deff} gives
\(a_{\rm LO}(t)=a_{{\rm LO},0}\), \(a_{\rm A}(t)=a_{{\rm A},0}\), and
\(\Delta_{\rm eff}(t)=\Delta_0\) for all \(t\). Hence,~\eqref{eq:Q} gives
\(Q(t)=Q_0\). Moreover, under the far-detuning case, the correction $\eta(t)$
is negligible by~\eqref{eq:E_bound}. Moreover, $|\Delta_0|\leq\sqrt{Q_0/C_1}$ ensures that
the broadening contribution to the gain is negligible.
Consequently, substituting these constants into
\eqref{eq:bias}, \eqref{eq:rho21_dress},
and~\eqref{eq:gain_dress}, we complete this proof of \eqref{eq:Pi0g0}.
\end{IEEEproof}
 
Theorem~\ref{thm:lock} shows that, with fixed lasers and
constant within-chirp transmit power, the bias and gain remain
constant at $\Pi_0$ and $g_0$, respectively. Consequently,
the bias is confined to the zero range
bin and removed by DC blocking, while the
gain-induced leakage in Remark~\ref{rem:leakage} is eliminated,
namely \(\hat g(\zeta)=g_0\mathcal W(\zeta)\), leaving each beat tone with
only the Dirichlet kernel of the observation window centered at its beat bin
\(n_{k,m}\). In the following,  we maximize
the receiver gain by designing $\Delta_0$, $a_{{\rm LO},0}$, and $a_{{\rm A},0}$.
 
\begin{proposition}[Optimal working point and LO dressing]\label{prop:opt}
Define \(\bar \chi=\frac{L_{\rm cell}k_{\rm p}D_{\Omega}B_1}{C_1}\). For a given \(a_{{\rm A},0}\), the magnitude of the locked gain
\(\lvert g_0\rvert\) in~\eqref{eq:Pi0g0} is maximized by 
\begin{equation}\label{eq:Delta_opt}
 \lvert\Delta_0^{\star}\rvert=\sqrt{\frac{x^{\star}Q_0}{C_1}},\quad
 x^{\star}=\frac{\sqrt{(1+\bar \chi)^2+3}-(1+\bar \chi)}{3}
\end{equation}
and
\begin{equation}\label{eq:aLO_opt}
 a_{{\rm LO},0}^{\star}=\sqrt{1+a_{{\rm A},0}^2
 +\frac{\Omega_{\rm c}^2}{\Omega_{\rm p}^2}}.
\end{equation}
Therefore, the optimal atomic gain is expressed as 
\begin{equation}\label{eq:g_opt}
\begin{split}
     \lvert g_0^{\star}\rvert=\frac{2\sqrt{x^{\star}}}{(1+x^{\star})^2}
 e^{-\frac{\bar \chi x^{\star}}{1+x^{\star}}}
 \frac{P_{\rm in}L_{\rm cell}k_{\rm p}D_{\Omega}B_1}
 {\sqrt{C_1}\,a_{{\rm LO},0}^{\star}\Omega_{\rm p}^2}.
\end{split}
\end{equation}
\end{proposition}

\begin{IEEEproof}
Define \(x=\frac{C_1\Delta_0^2}{Q_0}\). Using~\eqref{eq:Pi0g0}, we have
\(\lvert g_0\rvert=\frac{2P_{\rm in}L_{\rm cell}k_{\rm p}D_{\Omega}B_1
a_{{\rm LO},0}}{\sqrt{C_1Q_0}}\,h(x)\), where
\(h(x)=\frac{\sqrt{x}}{(1+x)^2}e^{-\frac{\bar \chi x}{1+x}}\). Since \(h(0)=0\) and
\(h(x)\to0\) with \(x\to\infty\), we set
\(\frac{{\rm d}\ln h(x)}{{\rm d}x}=\frac{1}{2x}-\frac{2}{1+x}
-\frac{\bar \chi}{(1+x)^2}\) to zero and have \(3x^2+2(1+\bar \chi)x-1=0\). Therefore,  \(\lvert g_0\rvert\) can be maximized by the
unique positive root \(x^{\star}\) in~\eqref{eq:Delta_opt}.
Moreover, since \(\bar \chi\) and \(x^{\star}\) are independent of
\(a_{{\rm LO},0}\), maximizing \(\lvert g_0\rvert\) over \(a_{{\rm LO},0}\) is equivalent to maximizing \(\frac{a_{{\rm LO},0}}{\sqrt{Q_0}}\). Setting the derivative of \(\frac{a_{{\rm LO},0}}{\sqrt{Q_0}}\) with respect to \(a_{{\rm LO},0}\) to zero, we obtain~\eqref{eq:aLO_opt}. Finally, we complete this proof by substituting \(\sqrt{Q_0}=a_{{\rm LO},0}^{\star2}\Omega_{\rm p}^2\)
into \(g_0\) in \eqref{eq:Pi0g0}.
\end{IEEEproof}
 
Proposition~\ref{prop:opt} provides a closed-form optimal design.
The optimal detuning in~\eqref{eq:Delta_opt} balances the
transparency-window slope against the transmitted probe power.
The optimal LO dressing in~\eqref{eq:aLO_opt} balances the
increased gain against dressing-induced broadening.
At the optimum,
$a_{{\rm LO},0}^{\star2}\Omega_{\rm p}^2
=\Omega_{\rm c}^2+(1+a_{{\rm A},0}^2)\Omega_{\rm p}^2$,
indicating that a weaker probe requires stronger LO dressing.
In contrast, AUX dressing broadens the window without directly
enhancing the gain, favoring a small $a_{{\rm A},0}$.
By~\eqref{eq:lock}, reducing $a_{{\rm A},0}$ increases the
AUX detuning,
$|\varepsilon_5(t)|\approx \frac{a_{{\rm LO},0}^2}{a_{{\rm A},0}^2}|\varepsilon_4(t)|$,
but requires a stronger AUX microwave field,
$\Omega_{\rm A}(t)\approx\frac{2a_{{\rm LO},0}^2}{a_{{\rm A},0}}|\varepsilon_4(t)|$,
revealing a tradeoff between gain optimization and AUX driving power.

\section{Performance Analysis}\label{sec:perf}

 In this section, we evaluate the resulting
sensing performance, including the output SNR, the CRB of the range and Doppler estimation, and the impacts of the system parameters.

\subsection{Output SNR}\label{sec:snr}

Following~\cite{Cuisensing}, we consider two noise sources, namely the external noise (ETN) and internal noise (ITN). Since the amplifier scales the signal and noise equally, we characterize both at the photocurrent level, where the bias and gain are $\mathcal R_{\rm ph}\Pi(t)$ and $\mathcal R_{\rm ph}g(t)$, respectively. ETN enters the atomic receiver together with the echoes and is converted through the atomic gain, whereas ITN arises from
shot noise associated with the bias photocurrent. Their power spectral densities are given by \cite{Cuisensing}
\begin{equation}
\begin{aligned}
\sigma^2_{\rm ETN}(t)
&=\left(\frac{\mu_{34}}{\hbar}\right)^2
\mathcal R_{\rm ph}^2g^2(t)\langle E_{\rm I}^2\rangle,\\
\sigma^2_{\rm ITN}(t)
&=q\mathcal R_{\rm ph}\Pi(t),
\end{aligned}
\label{eq:noise}
\end{equation}
where
$\langle E_{\rm I}^2\rangle
=\frac{\hbar\omega_{\rm RF}^3}{\pi\epsilon_0c^3}
(2n_{\rm th}+1)$
denotes the electric-field noise spectral density,
$n_{\rm th}=
[\exp(\hbar\omega_{\rm RF}/(k_{\rm B}T_{\rm E}))-1]^{-1}$
is the mean thermal photon occupation number, and
$k_{\rm B}$ and $T_{\rm E}$ denote the Boltzmann constant
and ambient temperature, respectively.

For the locked receiver, Theorem~\ref{thm:lock} gives \(\Pi(t)=\Pi_0\) and
\(g(t)=g_0\), leading to constant spectral densities. Following~\cite[Eq.~(50)]{Cuisensing}, the SNR of target \(k\) over the observation time \(MT\) of one frame is
\begin{equation}\label{eq:snr}
 \mathrm{SNR}_k
 =\frac{\big(\frac{\mu_{34}}{\hbar}\big)^2\mathcal R_{\rm ph}^2g_0^2
 \lvert\beta_k\rvert^2MT}{\sigma^2_{\rm ETN}+\sigma^2_{\rm ITN}}
 =\frac{\mathrm{SNR}_{{\rm ETN},k}\,\mathrm{SNR}_{{\rm ITN},k}}
 {\mathrm{SNR}_{{\rm ETN},k}+\mathrm{SNR}_{{\rm ITN},k}},
\end{equation}
where \(\lvert\beta_k\rvert^2=\sum_m\lvert\beta_kx_l[m]\rvert^2\) is the field
intensity of the echo of target \(k\), and
\begin{equation}\label{eq:snr_etn_itn}
\begin{aligned}
 \mathrm{SNR}_{{\rm ETN},k}&=\frac{MT}{\langle E_{\rm I}^2\rangle}\,
\frac{2Z_0P_{\rm tx}G_{\rm tx}}{A_{\rm c}}
\lvert h_k\rvert^2,\\
 \mathrm{SNR}_{{\rm ITN},k}&=\Big(\frac{\mu_{34}}{\hbar}\Big)^2
 \frac{\mathcal R_{\rm ph}g_0^2\,MT}{q\Pi_0}\,\frac{2Z_0P_{\rm tx}G_{\rm tx}}{A_{\rm c}}
\lvert h_k\rvert^2.
\end{aligned}
\end{equation}
Since ETN and the echoes undergo the same atomic conversion,
$\mathrm{SNR}_{{\rm ETN},k}$ is independent of the receiver
design and provides an upper bound on $\mathrm{SNR}_k$.
In contrast, $\mathrm{SNR}_{{\rm ITN},k}$ scales with
$\frac{g_0^2}{\Pi_0}$ and can be improved through receiver optimization. Moreover, note that both SNR terms remain constant throughout the chirp under
the proposed locking design, unlike the time-varying case
in~\cite{Cuisensing}.

\subsection{CRB of Delay and Doppler Estimation}\label{sec:crb}

We analyze the CRB of the delay and Doppler estimation
based on the samples \(V_l[i]\) in~\eqref{eq:samples}, where
\(\Pi[i]=\Pi_0\) and \(g[i]=g_0\) by Theorem~\ref{thm:lock}. Since the range
and the velocity of target \(k\) are determined by \(\tau_k\) and \(\nu_k\), we
estimate \(\boldsymbol\theta_k=[\tau_k,\nu_k]^{\rm T}\). Moreover, we assume
that the transmitted symbols are known at the receiver and that the targets
are resolvable.

By~\eqref{eq:envelope} and~\eqref{eq:beat}, the phase of the beat tone from
target \(k\) on subcarrier \(m\) at sample \(i\) of chirp \(l\) is
\(2\pi(f_{k,m}iT/N_{\rm FFT}+\psi_{k,m}+\nu_klT_{\rm r})\), whose gradient with
respect to \(\boldsymbol\theta_k\) is
\(2\pi[\alpha\tau_k-\alpha iT/N_{\rm FFT}-m\Delta f,\,
iT/N_{\rm FFT}+lT_{\rm r}]^{\rm T}\). Since the beat bins satisfy \(n_{k,m}=f_{k,m}T\gg1\), the double-frequency
terms average out over each chirp. Moreover, the zero-mean
and uncorrelated subcarrier symbols eliminate the cross-subcarrier terms in the
symbol-averaged Fisher Information Matrix (FIM), while the echo amplitude decouples from its phase. By
treating the unknown common phase
\(\angle\beta_k-2\pi f_{\rm c}(\tau_k-\tau_{\rm L})\) as a nuisance parameter
and eliminating it via the Schur complement~\cite{steven1993fundamentals}, we obtain the
symbol-averaged FIM as
\begin{equation}
\mathbf J(\boldsymbol\theta_k)
=2\pi^2\,\mathrm{SNR}_k\,
\mathrm{Cov}\left\{
\begin{bmatrix}
-\alpha\frac{iT}{N_{\rm FFT}}-m\Delta f\\[1mm]
\frac{iT}{N_{\rm FFT}}+lT_{\rm r}
\end{bmatrix}
\right\},
\label{eq:fim}
\end{equation}
where \(\mathrm{SNR}_k\) is given in~\eqref{eq:snr}, and
\(\mathrm{Cov}\{\cdot\}\) denotes the covariance over the samples \(i\), the
chirps \(l\), and the subcarriers \(m\), with the subcarriers weighted by
\(\lvert x_l[m]\rvert^2\). Since the common phase is unknown,~\eqref{eq:fim}
involves the covariances rather than the second moments of the phase gradient,
which removes the delay information associated with the carrier phase
\(2\pi f_{\rm c}\tau_k\). Therefore, the delay and Doppler estimation accuracies depend on
the spread of the instantaneous frequencies and of the observation times,
respectively.

\begin{proposition}[CRB of delay and Doppler]
\label{prop:crb}
For equal-power subcarriers with
$|x_l[m]|^2=\frac{1}{N_{\rm s}}$, the CRBs of
$\tau_k$ and $\nu_k$ are approximately given by
\begin{equation}
\mathrm{CRB}(\tau_k)\approx\frac{6\bar T^2}{\pi^2\mathrm{SNR}_k\Upsilon},\quad
\mathrm{CRB}(\nu_k)\approx\frac{6\bar B^2}{\pi^2\mathrm{SNR}_k\Upsilon},
\label{eq:crb}
\end{equation}
where \(\bar B^2=B^2+(N_{\rm s}^2-1)\Delta f^2\) and
\(\bar T^2=T^2+T_{\rm r}^2(M^2-1)\) are the frequency and time spreads of the
frame, and \(\Upsilon=\bar B^2\bar T^2-B^2T^2\).
\end{proposition}

\begin{IEEEproof}
For equal-power subcarriers, the covariance
in~\eqref{eq:fim} is evaluated over independent,
uniformly distributed indices
$i\in\{0,\ldots,N_{\rm FFT}-1\}$,
$m\in\{0,\ldots,N_{\rm s}-1\}$, and
$l\in\{0,\ldots,M-1\}$.
Their variances are given by
\begin{equation}
    \begin{aligned}
\mathrm{Var}\left\{\frac{iT}{N_{\rm FFT}}\right\}
&=\frac{T^2}{12}
\left(1-\frac{1}{N_{\rm FFT}^2}\right)
\approx\frac{T^2}{12},\\
\mathrm{Var}\{m\}
&=\frac{N_{\rm s}^2-1}{12},\quad
\mathrm{Var}\{l\}=\frac{M^2-1}{12},
\end{aligned}
\end{equation}
where the approximation holds for sufficiently large
$N_{\rm FFT}$. Using $B=\alpha T$, the covariance
between the two entries in~\eqref{eq:fim} is
approximately $\frac{-BT}{12}$.
Substituting these moments into~\eqref{eq:fim},
the FIM can be expressed as
\begin{equation}
    \mathbf J(\boldsymbol\theta_k)\approx\frac{\pi^2\mathrm{SNR}_k}{6}
\begin{bmatrix}
\bar B^2 & -BT\\
-BT & \bar T^2
\end{bmatrix}.
\end{equation}
We complete this proof by using the determinant and inverse  of \(\mathbf J(\boldsymbol\theta_k)\).
\end{IEEEproof}

For a sufficiently large number of chirps, the delay--Doppler
coupling becomes negligible. The delay accuracy is then primarily
determined by the swept bandwidth, whereas the Doppler accuracy
depends on the frame duration. Since the proposed locking design
maintains a constant SNR throughout the chirp, increasing the
swept bandwidth reduces the delay CRB approximately quadratically,
at the cost of stronger LO and AUX driving.
In contrast, the time-varying gain in~\cite{Cuisensing}
introduces nonuniform sensing information and spectral leakage,
requiring bias- and gain-aware estimation to approach the CRB.
By maintaining constant gain and stationary noise, the proposed
design eliminates receiver-induced leakage and enables 2D-FFT peak detection followed by a local maximum-likelihood
refinement of the off-grid delay and Doppler to approach the CRB at high SNR.

\subsection{Impacts of LO Chirp and AUX on Sensing}\label{sec:impact_sensing}

We first investigate the effects of the LO chirp and the AUX field on the conversion
gain \(\lvert g_0\rvert\) and on the CRBs in~\eqref{eq:crb}. Let
\(\varepsilon_4^{\max}=\max_t\lvert\varepsilon_4(t)\rvert
\approx\min_t\lvert\varepsilon_4(t)\rvert+2\pi B\), \(\Omega_{\rm LO}^{\max}\) and \(\Omega_{\rm A}^{\max}\) denote the maximum Rabi frequencies of the LO and AUX, respectively. As discussed in Section~\ref{sec:lock}, the LO light shift
dominates under
$a_{{\rm LO},0}^2|\varepsilon_4(t)|
\gg|\Delta_0-\varepsilon_3(t)|$,
ensuring that $\varepsilon_5(t)$ does not cross zero. Then,
\(\Omega_{\rm LO}(t)\leq\Omega_{\rm LO}^{\max}\) and
\(\Omega_{\rm A}(t)\leq\Omega_{\rm A}^{\max}\) leads to \( a_{{\rm LO},0}\leq\frac{\Omega_{\rm LO}^{\max}}{2\varepsilon_4^{\max}}\) and \( a_{{\rm A},0}\geq\frac{2a_{{\rm LO},0}^2\varepsilon_4^{\max}}
 {\Omega_{\rm A}^{\max}}\), which means the LO amplitude limits the LO dressing from above, whereas the AUX amplitude limits the AUX dressing from below.

\begin{proposition}[Upper bound of the locked gain]\label{prop:gain_fact}
At the working point given by~\eqref{eq:Delta_opt}, the locked gain satisfies
\begin{equation}\label{eq:g_fact}
 \lvert g_0\rvert=\bar g\,\kappa_{\rm LO}\,\kappa_{\rm A}
 \leq\bar g\,\min\bigg\{1,\frac{2a_{{\rm LO},0}}{\bar a}\bigg\},
\end{equation}
where
\begin{equation}\label{eq:kappa}
 \kappa_{\rm LO}=\frac{2\bar a\,a_{{\rm LO},0}}{\bar a^2+a_{{\rm LO},0}^2},\quad
 \kappa_{\rm A}=\frac{\bar a^2+a_{{\rm LO},0}^2}
 {\bar a^2+a_{{\rm LO},0}^2+a_{{\rm A},0}^2}.
\end{equation}
$\bar a=\sqrt{1+\Omega_{\rm c}^2/\Omega_{\rm p}^2}$
and $\bar g$ denotes the reference gain obtained
from~\eqref{eq:g_opt} by replacing
$a_{{\rm LO},0}^{\star}$ with $\bar a$. The LO factor reaches $\kappa_{\rm LO}=1$ only at
$a_{{\rm LO},0}=\bar a$, whereas the AUX factor
satisfies \(\kappa_{\rm A}\approx1\) whenever
\(a_{{\rm A},0}^2\ll\bar a^2+a_{{\rm LO},0}^2\).
Consequently, the ITN-limited SNR is upper bounded as
\begin{equation}\label{SNRup}
    \mathrm{SNR}_{{\rm ITN},k}
=\kappa_{\rm LO}^2\kappa_{\rm A}^2
\overline{\mathrm{SNR}}_{{\rm ITN},k}
\leq\overline{\mathrm{SNR}}_{{\rm ITN},k},
\end{equation}
where $\overline{\mathrm{SNR}}_{{\rm ITN},k}$
is obtained from~\eqref{eq:snr_etn_itn} with \(g_0 = \bar g\).
\end{proposition}

\begin{IEEEproof}
By using Proposition~\ref{prop:opt}, we have \(\lvert g_0\rvert=\frac{2P_{\rm in}L_{\rm cell}k_{\rm p}D_{\Omega}B_1
a_{{\rm LO},0}}{\sqrt{C_1Q_0}}h(x^{\star})\). Then, using $\sqrt{Q_0}=\frac{\Omega_{\rm p}^2}{2}
(\bar a^2+a_{{\rm LO},0}^2+a_{{\rm A},0}^2)$, we prove the equality of~\eqref{eq:g_fact}.
Since $\kappa_{\rm A}\leq1$ and
$\bar a^2+a_{{\rm LO},0}^2
\geq\max\{2\bar a a_{{\rm LO},0},\bar a^2\}$,
the upper bound of locked gain is proved. Finally, since $\Pi_0$ depends on the dressing ratios only through $\frac{C_1\Delta_0^2}{Q_0}=x^{\star}$,
$\mathrm{SNR}_{{\rm ITN},k}$ scales with $g_0^2$, we prove the upper bound of ITN-limited SNR.
\end{IEEEproof}

For \(M\gg1\) and \(BT\gg N_{\rm s}\),~\eqref{eq:crb} reduces to
\(\mathrm{CRB}(\tau_k)\approx\frac{6}{\pi^2B^2\mathrm{SNR}_k}\) and
\(\mathrm{CRB}(\nu_k)\approx\frac{6}{\pi^2T_{\rm r}^2(M^2-1)\mathrm{SNR}_k}\).
Consequently, the LO and AUX affect both CRBs through \(\kappa_{\rm LO}\kappa_{\rm A}\) in the ITN term, whereas the LO additionally affects \(\mathrm{CRB}(\tau_k)\) through the sweep span \(B\).

\textit{Impact of LO.} The peak LO amplitude limits the dressing strength available across the sweep.
For a given \(a_{{\rm A},0}\), the gain
in~\eqref{eq:g_fact} peaks at \(a_{{\rm LO},0}^{\star}\). Considering the maximum Rabi frequency of LO, the optimal feasible dressing is
\(a_{{\rm LO},0}=\min\{a_{{\rm LO},0}^{\star},
\frac{\Omega_{\rm LO}^{\max}}{2\varepsilon_4^{\max}}\}\). When
\(\varepsilon_4^{\max}\leq\frac{\Omega_{\rm LO}^{\max}}{2a_{{\rm LO},0}^{\star}}\),
widening the sweep preserves the optimal gain and reduces
the delay CRB as $B^{-2}$ without affecting the Doppler CRB
at fixed frame timing. Beyond this regime, the admissible LO dressing decreases as
\(\frac{1}{\varepsilon_4^{\max}}\), resulting in a reduced \(\mathrm{SNR}_{{\rm ITN},k}\). Using
\(\varepsilon_4^{\max}>2\pi B\) and
Proposition~\ref{prop:gain_fact}, the delay CRB
satisfies
\begin{equation}\label{eq:crb_floor}
\begin{aligned}
 \mathrm{CRB}(\tau_k)\gtrsim{}&\frac{6}{\pi^2B^2\,\mathrm{SNR}_{{\rm ETN},k}}+\frac{24\,\bar a^2}{\overline{\mathrm{SNR}}_{{\rm ITN},k}
 \big(\Omega_{\rm LO}^{\max}\big)^2}.
\end{aligned}
\end{equation}
As can be seen, the peak LO amplitude imposes a floor on the ITN contribution to the delay CRB, while the ITN contribution to the Doppler CRB is lower bounded in
proportion to \(\big(\bar a\,\varepsilon_4^{\max}/\Omega_{\rm LO}^{\max}\big)^2\).
In the amplitude-limited regime, widening the sweep therefore introduces a
delay-Doppler accuracy tradeoff.

\textit{Impact of AUX.}  The AUX affects sensing only by broadening the transparency window, and this effect vanishes as the AUX strengthens. Specifically, since the echoes drive only
\(|3\rangle\rightarrow|4\rangle\), the AUX dressing contributes to the window
width through \(a_{{\rm A},0}\), rather than  the light-shift rate. At the optimal LO dressing \(a_{{\rm LO},0}^{\star}\), the AUX-induced
broadening reduces the ITN-limited SNR to
\(\overline{\mathrm{SNR}}_{{\rm ITN},k}/(1+a_{{\rm A},0}^2/\bar a^2)\), thereby increasing both CRBs. However, a stronger AUX supplies the required light
shift with a smaller \(a_{{\rm A},0}\), and thus \(\kappa_{\rm A}\) approaches
unity and the penalty becomes negligible once
\(a_{{\rm A},0}^2\ll\bar a^2+a_{{\rm LO},0}^2\). This condition is readily met
by a moderate AUX for an amplitude-limited LO with \(a_{{\rm LO},0}\ll\bar a\),
whereas operation near \(a_{{\rm LO},0}^{\star}\) requires a stronger AUX.

In the four-level receiver of~\cite{Cuisensing}, the LO controls both the
dressing strength and the operating point, with
\(a_{\rm LO}^2(t)\lvert\varepsilon_4(t)\rvert=\lvert\Delta_{\rm eff}(t)\rvert\)
for \(\varepsilon_3(t)=0\). Maintaining operation within the
transparency window requires
\(a_{\rm LO}(t)\leq\big[\sqrt{Q(t)/C_1}/\lvert\varepsilon_4(t)\rvert\big]^{1/2}\). Meanwhile, since \(h(x)\leq h(x^{\star})\),~\eqref{eq:g_fact} also bounds the four-level gain as \(\lvert g(t)\rvert\leq\frac{2\bar g\,a_{\rm LO}(t)}{\bar a}\). Consequently, when the LO
detuning greatly exceeds the window half-width, the gain remains far below
\(\bar g\) regardless of the LO power. In contrast, the AUX compensates the
LO-induced light shift, allowing the dressing strength and the operating point
to be optimized independently and held constant throughout the chirp. As a result, with
sufficient LO and AUX driving, the locked gain based on our configuration can approach
\(\bar g\).

\section{Simulation Results}
In this section, we first characterize the benefits of the five-level configuration and then evaluate its performance in sensing.

\begin{table*}[t]
\centering
\caption{Simulation Parameters}
\label{tab:simulation_parameters}
\footnotesize
\renewcommand{\arraystretch}{1.25}
\setlength{\tabcolsep}{3pt}
\begin{tabularx}{\textwidth}{@{}>{\raggedright\arraybackslash\hsize=1.15\hsize}X c
>{\raggedright\arraybackslash\hsize=0.9\hsize}X c
>{\raggedright\arraybackslash\hsize=0.95\hsize}X c@{}}
\rowcolor{tableblue}
\multicolumn{2}{c}{\textbf{Atomic Parameters}} &
\multicolumn{2}{c}{\textbf{Cell, Readout, and Lock}} &
\multicolumn{2}{c}{\textbf{AFDM Waveform and Scene}}\\
\toprule
\textbf{Parameter} & \textbf{Value} &
\textbf{Parameter} & \textbf{Value} &
\textbf{Parameter} & \textbf{Value}\\
\midrule
\rowcolor{gray!4}
Probe Rabi frequency \(\Omega_{\rm p}/(2\pi)\) & \(1.69\)~MHz &
Cell length \(L_{\rm cell}\) & \(6.5\)~cm &
Carrier frequency \(f_{\rm c}\) & \(16.1\)~GHz\\
Coupling Rabi frequency \(\Omega_{\rm c}/(2\pi)\) & \(15\)~MHz &
Probe power \(P_{\rm in}\) & \(1~\mu\)W &
Sweep span \(B\) & \(150\)~MHz\\
\rowcolor{gray!4}
Laser detunings \(\Delta_{\rm p},\Delta_{\rm c}\) & \(0\)~MHz &
Coupling power \(P_{\rm c}\) & \(3\)~W &
Useful duration \(T\) & \(1\)~ms\\
Decay rate \(\gamma_2/(2\pi)\) & \(6.07\)~MHz &
Beam diameters \(d_{\rm p}=d_{\rm c}\) & \(0.7\)~mm &
CPP duration \(T_{\rm cp}\) & \(10~\mu\)s\\
\rowcolor{gray!4}
Probe wavelength \(\lambda_{\rm p}\) & \(780.241\)~nm &
PD responsivity \(\mathcal R_{\rm ph}\) & \(0.5494\)~A/W &
Chirps per frame \(M\) & \(64\)\\
Coupling wavelength \(\lambda_{\rm co}\) & \(\approx480.1\)~nm &
Power gain \(G_{\rm ph}\) & \(30\)~dB &
Active subcarriers \(N_{\rm s}\) & \(16\)\\
\rowcolor{gray!4}
Probe dipole \(\lvert\mu_{12}\rvert\) & \(2.989\,ea_0\) &
Ambient temperature \(T_{\rm E}\) & \(300\)~K &
FFT size \(N_{\rm FFT}\) & \(512\)\\
Coupling dipole \(\lvert\mu_{23}\rvert\) & \(0.01527\,ea_0\) &
LO detuning \(\min_t\lvert\varepsilon_4(t)\rvert/(2\pi)\) & \(70\)~MHz &
Transmit power \(P_{\rm tx}\) & \(20\)~dBm\\
\rowcolor{gray!4}
LO dipole \(\lvert\mu_{34}\rvert\) & \(2117.49\,ea_0\) &
LO dressing ratio \(a_{{\rm LO},0}\) & \(0.5\) &
Transmit antenna gain \(G_{\rm tx}\) & \(10\)~dBi\\
AUX dipole \(\lvert\mu_{35}\rvert\) & \(2176.08\,ea_0\) &
AUX dressing ratio \(a_{{\rm A},0}\) & \(0.5\) &
Radar cross section \(\sigma_k\) & \(10\)~dBsm\\
\rowcolor{gray!4}
Atomic temperature \(T_{\rm a}\) & \(100~\mu\)K &
LO delay \(\tau_{\rm L}\) & \(2~\mu\)s &
Target delays \(\tau_k\) & \(1.00,1.29,1.60~\mu\)s\\
Atomic density \(N_0\) & \(5.89\times10^{14}~\mathrm{m}^{-3}\) &
LPF cutoff \(f_{\rm LP}\) & \(200\)~kHz &
Target Doppler shifts \(\nu_k\) & \(315.78,-158.96,0\)~Hz\\
\bottomrule
\end{tabularx}
\vspace{-0.5cm}
\end{table*}

\begin{figure*}[h]
    \centering
    \includegraphics[width=.9\linewidth]{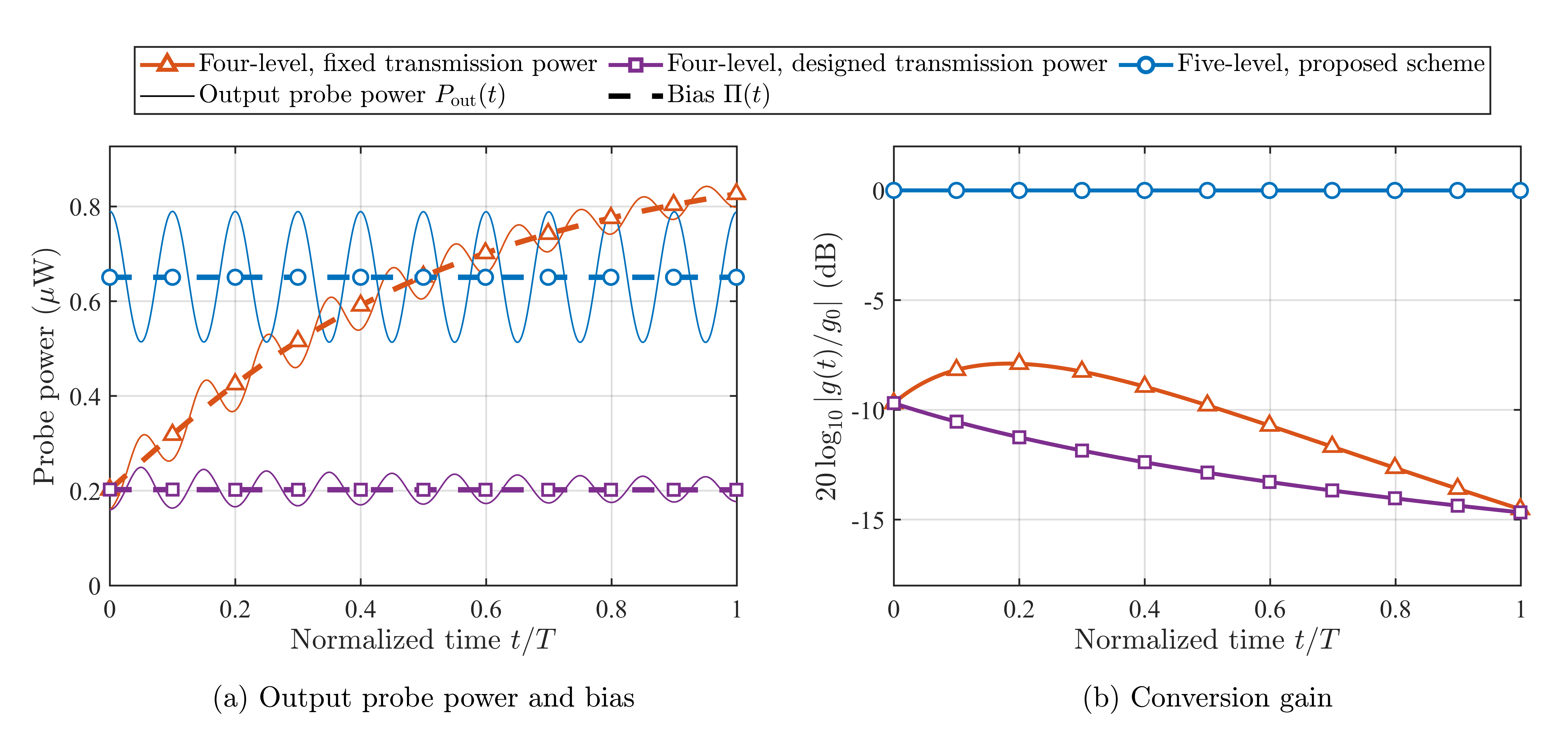}
    \caption{Output probe and atomic gain.}
    \label{biasandgain}
    \vspace{-0.5cm}
\end{figure*}

\begin{figure*}[h]
    \centering
    \includegraphics[width=.9\linewidth]{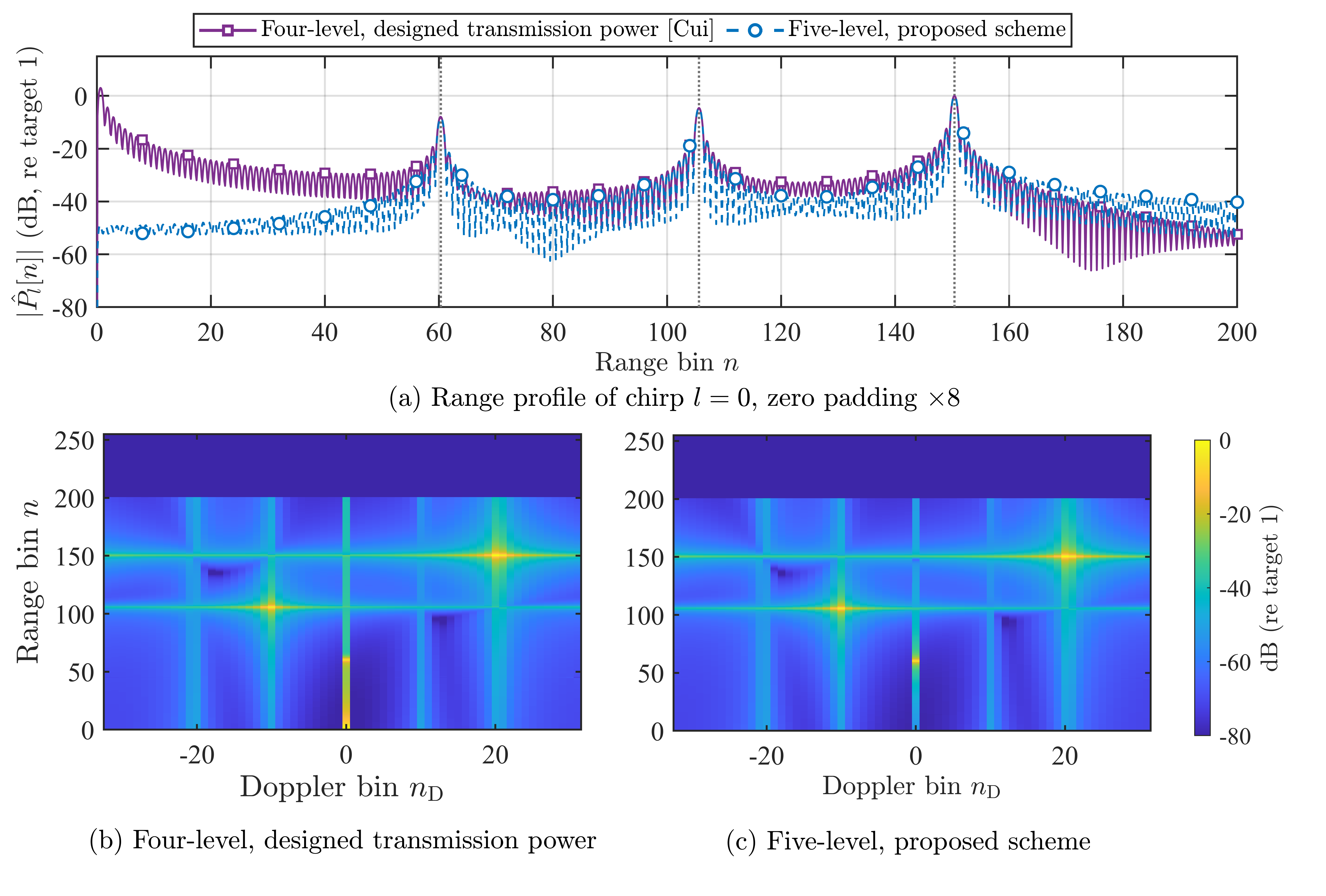}
    \caption{Range profiles and range-Doppler maps after DC blocking and low-pass filtering.}
    \label{fftfi}
    \vspace{-0.5cm}
\end{figure*}
\subsection{Parameter Settings}
We consider a laser-cooled $^{87}\mathrm{Rb}$ ensemble with
$|1\rangle=|5S_{1/2},F=2\rangle$,
$|2\rangle=|5P_{3/2},F'=3\rangle$,
$|3\rangle=|51D_{5/2},m_J=5/2\rangle$,
$|4\rangle=|52P_{3/2},m_J=3/2\rangle$, and
$|5\rangle=|50F_{7/2},m_J=7/2\rangle$.
Residual two-photon Doppler broadening is assumed negligible. Unless otherwise stated, the parameters
in Table~\ref{tab:simulation_parameters} are used.
The LO sweeps a bandwidth $B$ above the
$|3\rangle\leftrightarrow|4\rangle$ transition from the prescribed
minimum detuning. Locking follows~\eqref{eq:lock}, with
$|\Delta_0|=|\Delta_0^\star|$ from~\eqref{eq:Delta_opt}.
For comparison, the four-level receiver in~\cite{Cuisensing}
uses the same lasers and cell with
$\Omega_{\rm LO}^2(t)=k_1|\varepsilon_4(t)|$,
where $k_1$ is specified in~\cite[Theorem~1]{Cuisensing}.

\subsection{Bias and Gain}

Fig.~\ref{biasandgain} compares the output probe power, bias, and
conversion gain over one chirp. For the four-level receiver with
a fixed-power LO, the magnitude of the LO light shift
$\frac{\Omega_{\rm LO}^2}{4\varepsilon_4(t)}$ decreases as
$|\varepsilon_4(t)|$ increases, leading to the time-varying operating point, bias, and gain. By adopting the LO power law in~\cite{Cuisensing}, this light shift remains constant, yielding a nearly DC bias, but the gain remains time-varying as
$a_{\rm LO}(t)\propto|\varepsilon_4(t)|^{-1/2}$ decreases along
the chirp. In contrast, the proposed scheme uses the AUX light
shift to compensate for the varying LO light shift while holding
$\Delta_{\rm eff}(t)$, $a_{\rm LO}(t)$, and $a_{\rm A}(t)$
constant. This stabilizes the bias and gain, as shown in Theorem~\ref{thm:lock}. Furthermore, by decoupling the operating point from the LO dressing, the AUX field allows the optimal operation in Proposition~\ref{prop:opt} without weakening the dressing along the chirp.

Then, we evaluate the range profile and the range-Doppler maps after DC blocking and low-pass filtering in Fig.~\ref{fftfi}. The four-level receiver of~\cite{Cuisensing} exhibits a strong background that decays slowly across the range bins. Although its LO power profile reduces bias variation, the residual drift remains comparable to the weak echoes. The mismatch between the bias values at the chirp boundaries produces slowly decaying spectral leakage beyond DC. Because this drift repeats across chirps, the background concentrates at zero Doppler in Fig.~\ref{fftfi}(b), making static and near-zero-Doppler targets particularly vulnerable to masking. This behavior agrees with Remark~\ref{rem:leakage}. In contrast, the proposed range profile in Fig.~\ref{fftfi}(a) closely matches the ideal LTI response. The lock stabilizes both the bias and gain, allowing DC blocking to remove the bias while suppressing gain-induced leakage. The cleaner zero-Doppler column in
Fig.~\ref{fftfi}(c) makes the static target clearly
distinguishable and allows a standard 2D FFT to resolve all
targets without time-varying bias subtraction or gain-aware
estimation.

\subsection{Optimum Design}

\begin{figure}[t]
    \centering
    \includegraphics[width=1\linewidth]{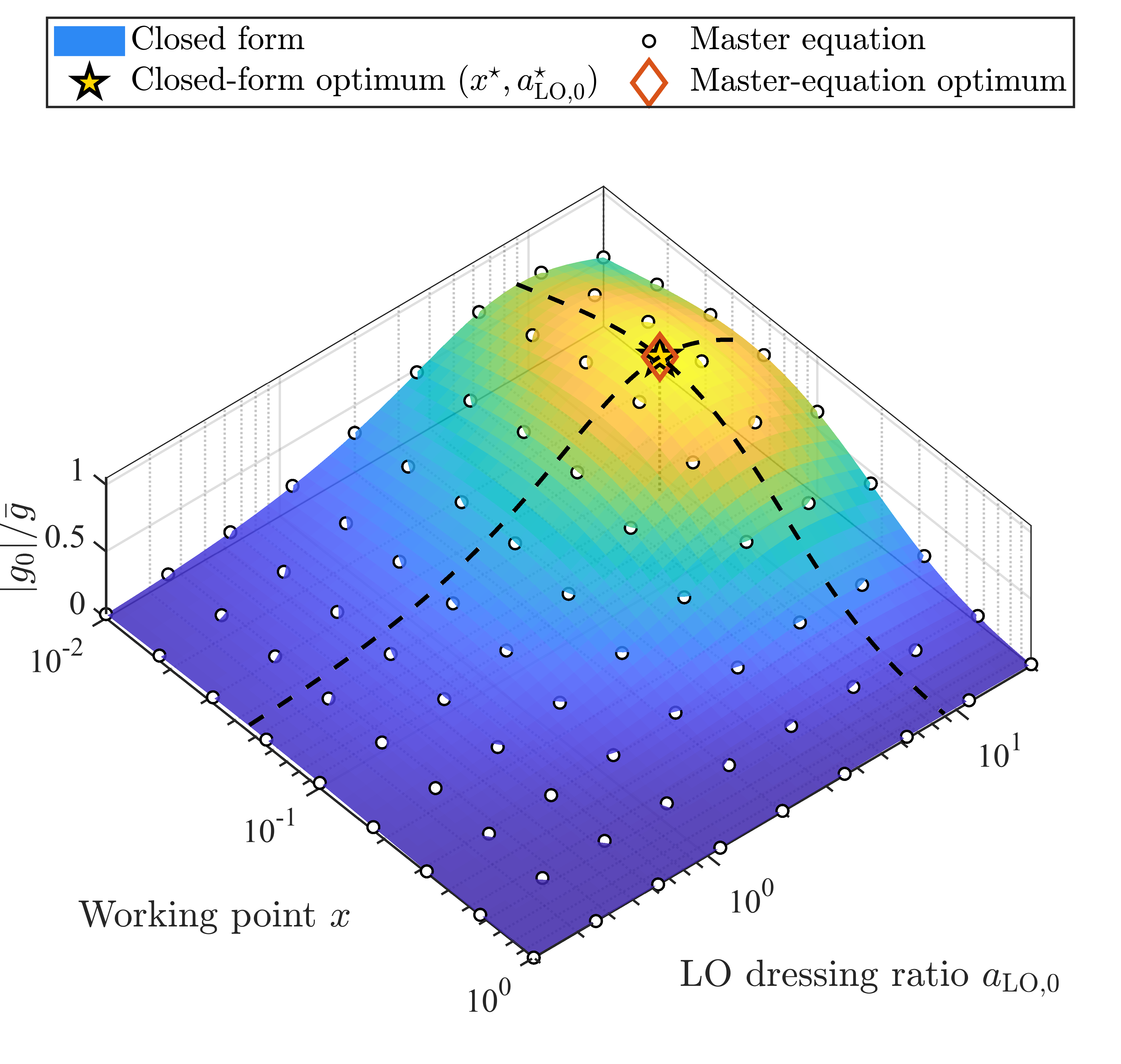}
    \caption{Optimal 5-level configuration at various working points and \(a_{{\rm LO},0}\).}
    \label{figoptdesign}
\end{figure}

\begin{figure}[t]
    \centering
    \includegraphics[width=1\linewidth]{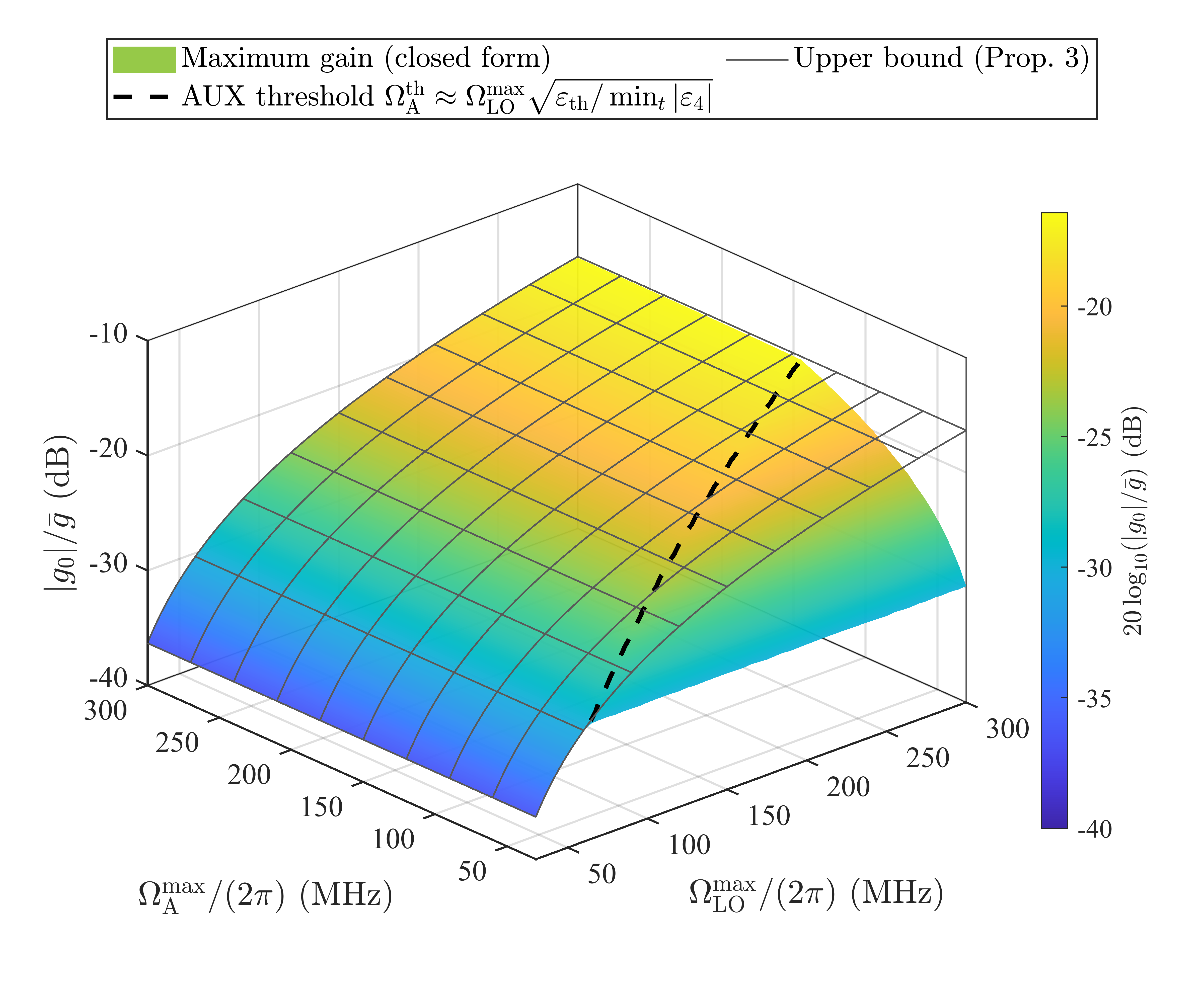}
    \caption{Maximum locked gain \(\frac{|g_0|}{\bar g}\) of the five-level receiver versus the peak LO and AUX Rabi frequencies.}
    \label{figLOAUX}
\end{figure}
In Fig.~\ref{figoptdesign}, we investigate the locked gains under various working points $x=\frac{C_1\Delta_0^2}{Q_0}$ and the LO dressing ratios $a_{{\rm LO},0}$. As observed, the maximum gain is achieved at $(x^{\star},a_{{\rm LO},0}^{\star})$, consistent with our Proposition~\ref{prop:opt}.  Physically, the gain is maximized by selecting the working point to balance the transparency-window slope against the transmitted probe power and the LO dressing to balance the increased light-shift rate against window broadening. Furthermore, compared with the coupled working point and LO dressing, the AUX field allows these two variables to be optimized
independently in closed form. This separated control enables the constant bias and gain, resulting in an LTI atomic conversion.

Considering the limited power of LO and AUX, we depict the maximum locked gain versus the peak LO and AUX Rabi frequencies, where the dressing ratios maximize $|g_0|$.  In Fig.~\ref{figLOAUX}, when the AUX is sufficiently strong, the gain grows approximately linearly with the peak LO amplitude and closely approaches the upper bound in Proposition~\ref{prop:gain_fact}, while further AUX strengthening provides little improvement. However, a weaker AUX requires stronger dressing for light-shift compensation, bringing $|5\rangle$ closer to resonance.
The far-detuning constraint therefore caps the LO dressing,
causing the gain to saturate despite increasing LO amplitude. Therefore, the AUX needs only enough amplitude to reach the
LO-limited regime, beyond which additional power provides
little benefit. In this regime, the normalized gain is
approximately $\frac{\Omega_{\rm LO}^{\max}}{\bar a\,\varepsilon_4^{\max}}$, explaining both the gap from $\bar g$ under the LO amplitude limit and the gain reduction as the sweep widens.

\subsection{Sensing Performance}

\begin{figure*}[t]
    \centering
    \includegraphics[width=.9\linewidth]{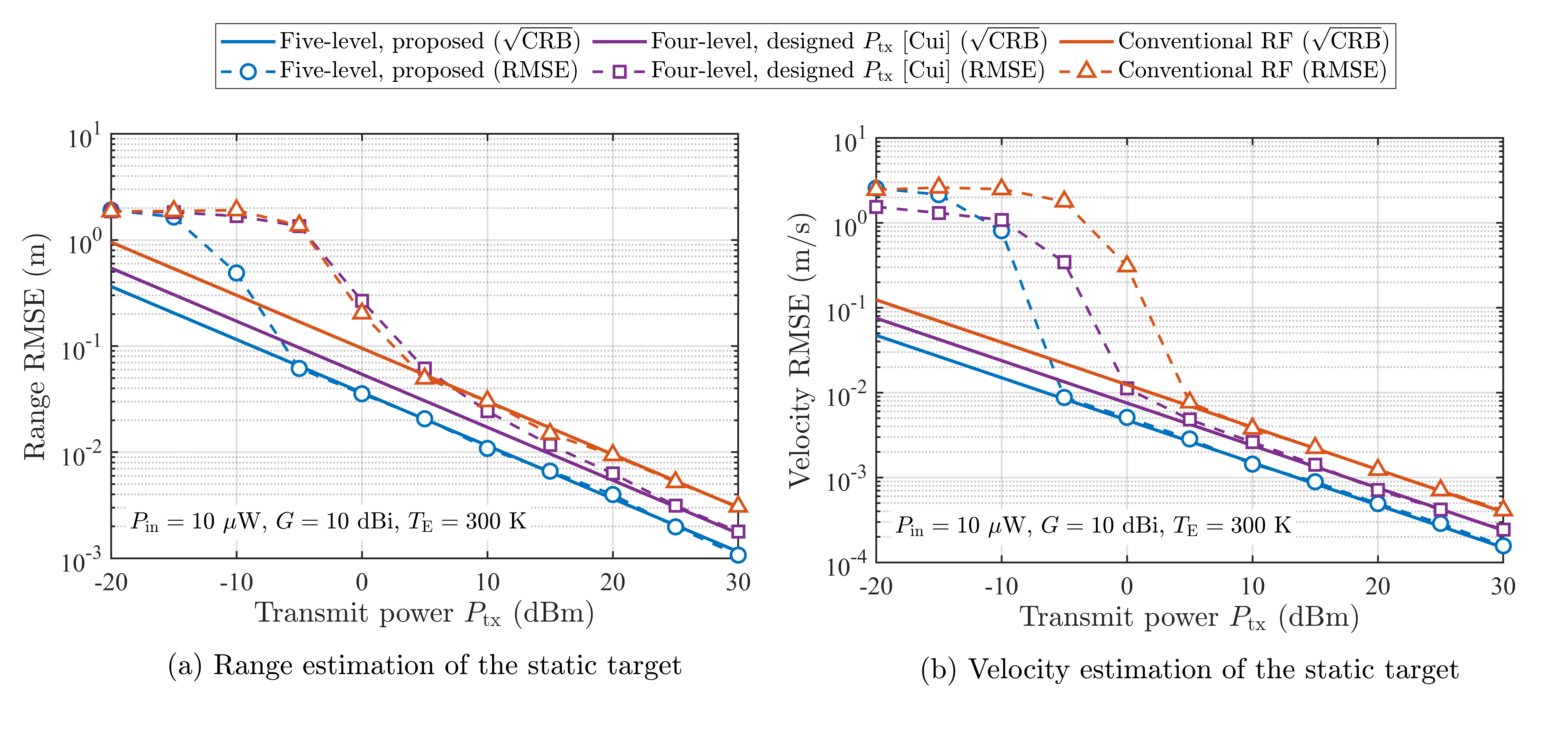}
    \caption{CRBs under various transmission powers.}
    \label{fig5CRB}
    \vspace{-0.5cm}
\end{figure*}

\begin{figure*}[t]
    \centering
    \includegraphics[width=.9\linewidth]{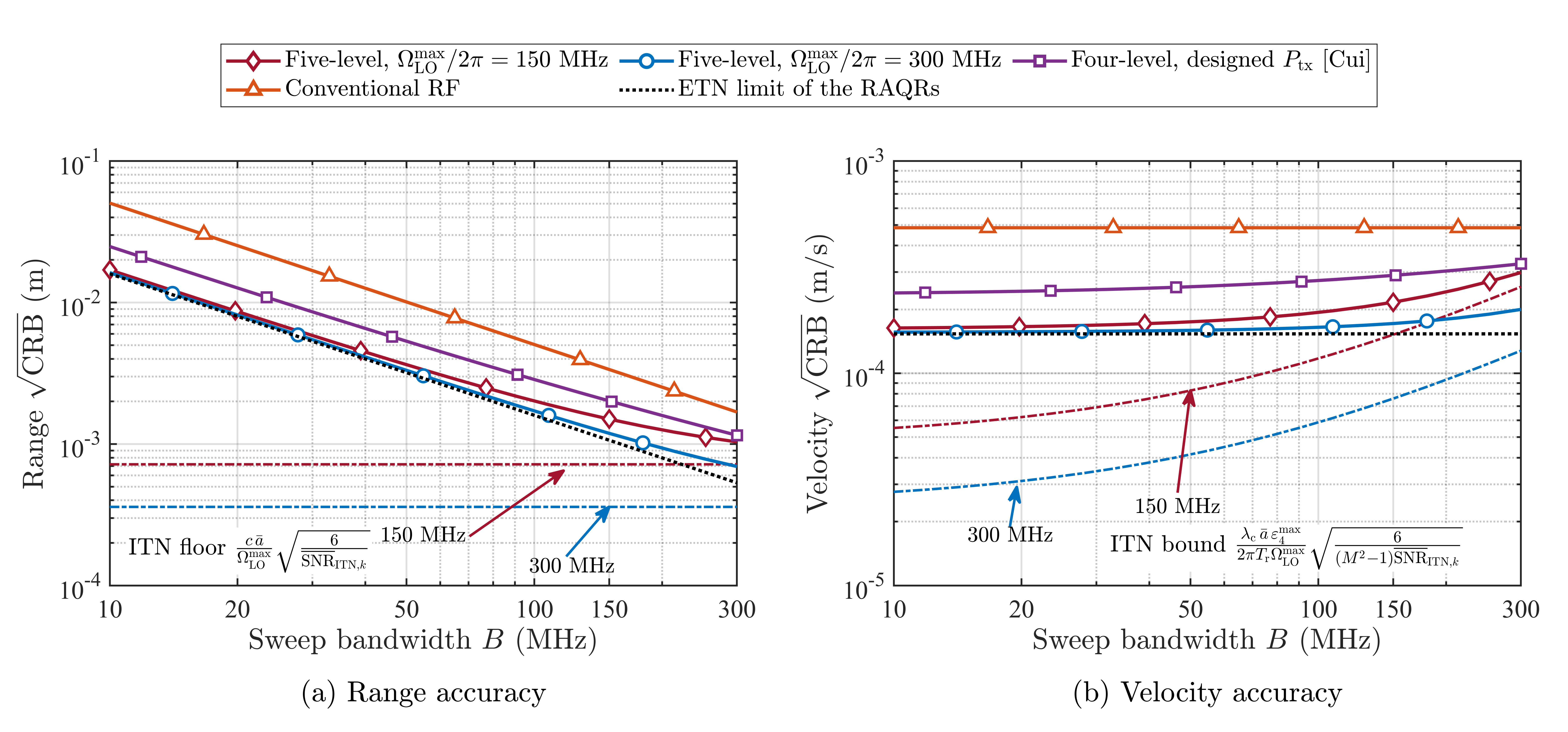}
    \caption{The impacts of sweep bandwidth on CRBs.}
    \label{fig6swb}
    \vspace{-0.5cm}
\end{figure*}

Fig.~\ref{fig5CRB} shows that the proposed five-level receiver
achieves lower range and velocity RMSEs than the four-level and RF benchmarks. This is attributed to the fact that the stabilized bias and gain suppress bias leakage and gain-induced range spreading, avoiding spectral leakage. Furthermore, a larger $g_0^2/\Pi_0$ also increases the ITN-limited SNR, while the vapor-cell gain provides an ETN advantage over the RF benchmark under the adopted noise model. Together, these improvements allow a 2D FFT to approach both CRBs in Proposition~\ref{prop:crb} at lower transmit
power, without time-varying bias subtraction or gain-aware
estimation.

Fig.~\ref{fig6swb} illustrates the sensing performance under various sweep bandwidths. As expected, our proposed scheme is superior to the 4-level scheme and RF benchmark. Furthermore, we note that widening the sweep initially reduces the range error bound as $B^{-1}$, but beyond a knee the bound approaches a floor while the velocity bound increases. This occurs because the peak LO amplitude limits
the dressing ratio to
$a_{{\rm LO},0}\leq\Omega_{\rm LO}^{\max}/(2\varepsilon_4^{\max})$, the
locked gain decreases as the sweep widens, reducing
$\mathrm{SNR}_{{\rm ITN},k}$ as
$(\varepsilon_4^{\max})^{-2}$.  Once ITN dominates, this SNR
loss offsets the bandwidth benefit for range estimation,
producing the floor predicted by~\eqref{eq:crb_floor}, while
directly degrading velocity estimation.

\section{Conclusions}
This paper developed a LTI five-level RAQR framework for AFDM sensing by using an AUX microwave field to compensate for chirp-induced time-varying atomic response.  Under far detuning, the five-level receiver was approximated by a dressed three-level EIT ladder, enabling a joint Stark-and-gain lock and stabilizing the bias and conversion gain. The AUX decoupled the operating
point from the LO dressing, and thus their optimal design was derived in closed-form. The analysis revealed that a sufficiently strong AUX allowed LO-limited operation, while increasing the sweep bandwidth beyond a threshold introduced a range-accuracy floor and a delay-Doppler tradeoff. Simulation results validated these findings and showed that the proposed receiver achieved a higher ITN-limited SNR and approached the CRBs with asimple FFT-based estimation. These results highlight the advantages of the five-level RAQR architecture for future high-resolution radar sensing applications.

\begin{appendices}

\section{Proof of  Proposition \ref{prop:three}}\label{Proofthree}
Let $\Xi(t)=a_{\rm LO}^2(t)+a_{\rm A}^2(t)$.
Using $\varepsilon_{\min}(t)$,  we have
\begin{equation}
    |E(t)|\le
\frac{\Omega_{\rm p}^4\Xi(t)|\Delta_{\rm eff}(t)|}
{2\varepsilon_{\min}(t)}
+\frac{\Omega_{\rm p}^4\Delta_{\rm eff}^2(t)}
{2\varepsilon_{\min}^2(t)}.
\end{equation}
Since $C_1\ge2\Omega_{\rm p}^2$ and
$Q(t)\ge\frac{[1+\Xi(t)]^2\Omega_{\rm p}^4}{4}$,
the denominator bounds can be written as
\begin{equation}
    \begin{aligned}
C_1\Delta_{\rm eff}^2(t)+Q(t)
&\ge2\sqrt{C_1Q(t)}\,|\Delta_{\rm eff}(t)|\\
&\ge\sqrt{2}[1+\Xi(t)]\Omega_{\rm p}^3
|\Delta_{\rm eff}(t)|,\\
C_1\Delta_{\rm eff}^2(t)+Q(t)
&\ge2\Omega_{\rm p}^2\Delta_{\rm eff}^2(t).
\end{aligned}
\end{equation}
Applying these bounds to the first and second residual terms, $\eta(t)$ is upper bounded by 
\[
|\eta(t)|\le
\frac{\Xi(t)}{1+\Xi(t)}
\frac{\Omega_{\rm p}}{2\sqrt{2}\,\varepsilon_{\min}(t)}
+\frac{\Omega_{\rm p}^2}{4\varepsilon_{\min}^2(t)}.
\]
Finally, removing $|4\rangle$ and $|5\rangle$ from
\eqref{eq:rho21_exact} gives the three-level response.
After substituting the effective parameters
in~\eqref{eq:dressed_par}, this response can be written as
\begin{equation}
    \begin{aligned}
 \Im\{\rho_{21}(t)\}=&-\frac{B_1\tilde\Delta^2(t)}
{C_1\tilde\Delta^2(t)
+\frac14[\tilde\Omega_{\rm c}^2(t)+\Omega_{\rm p}^2]^2}\\
=&
-\frac{B_1\Delta_{\rm eff}^2(t)}
{C_1\Delta_{\rm eff}^2(t)
+\frac14\big[\Omega_{\rm c}^2+
[1+\Xi(t)]\Omega_{\rm p}^2\big]^2}.
\end{aligned}
\end{equation}
Therefore, we complete this proof by setting \(\eta(t) = 0\) in \eqref{eq:rho21_eta}.

\end{appendices}

\bibliographystyle{IEEEtran}
\bibliography{myref}

\end{document}